\documentclass[11pt]{article}
\usepackage{times}
\usepackage{sidecap}
\usepackage{geometry}
\usepackage{floatrow}
\usepackage{float}
\usepackage[colorlinks=true, citecolor=blue, linkcolor=blue, urlcolor=blue]{hyperref}
\usepackage{graphics,graphicx}
\usepackage{subcaption}
\usepackage{epsf,wrapfig}
\usepackage{amssymb}
\usepackage{caption}
\usepackage{xcolor}
\usepackage{tcolorbox}
\usepackage{soul}
\usepackage{natbib,bm}
\usepackage{paralist}

\let\olditem\item
\renewenvironment{thebibliography}[1]{%
  \section*{\refname}
  \let\par\relax
  \renewcommand{\item}[1][]{\olditem[\textbullet]}%
  \inparaenum}{\endinparaenum}

\newcounter{aff}

\DeclareGraphicsExtensions{.jpg,.pdf,.png,.eps}

\bibpunct{(}{)}{;}{a}{}{,}

\definecolor{darkcyan}{rgb}{0.0, 0.0, 0.7}

\definecolor{boxcolor}{RGB}{193, 229, 255}
\definecolor{figboxcolor}{RGB}{193, 229, 255}
\definecolor{boxcolorgray}{RGB}{227, 227, 227}
\definecolor{darkcyan}{rgb}{0.0, 0.0, 0.7}

\sethlcolor{boxcolorgray}

\newcommand{\cii}{[C\,{\sc ii}]}

\newcommand{\ci}{[C\,{\sc i}]}

\newcommand{\oiiil}{[O\,{\sc iii}] 88\,$\mu{\rm m}$}
\newcommand{\ciil}{[C\,{\sc ii}] 158\,$\mu{\rm m}$}

\def\gtrsim{\lower 2pt \hbox{$\, \buildrel {\scriptstyle >}\over
{\scriptstyle \sim}\,$}}
\def\lesssim{\lower 2pt \hbox{$\, \buildrel {\scriptstyle <}\over
{\scriptstyle \sim}\,$}}

\def\approxlt{\lower.2em\hbox{$\buildrel < \over \sim$}}
\def\approxgt{\lower.2em\hbox{$\buildrel > \over \sim$}}

\def\be{\begin{equation}}
\def\ee{\end{equation}}
\def\bea{\begin{eqnarray*}}
\def\eea{\end{eqnarray*}}

\def\jref@jnl#1{{\rm#1}}
\def\aj{\jref@jnl{AJ}}			
\def\araa{\jref@jnl{ARA\&A}}		
\def\apj{\jref@jnl{ApJ}}			
\def\apjl{\jref@jnl{ApJ}}		
\def\apjs{\jref@jnl{ApJS}}		
\def\ao{\jref@jnl{Appl.~Opt.}}		
\def\apss{\jref@jnl{Ap\&SS}}		
\def\aap{\jref@jnl{A\&A}}		
\def\aapr{\jref@jnl{A\&A~Rev.}}		
\def\aaps{\jref@jnl{A\&AS}}		
\def\azh{\jref@jnl{AZh}}			
\def\baas{\jref@jnl{BAAS}}		
\def\jrasc{\jref@jnl{JRASC}}		
\def\memras{\jref@jnl{MmRAS}}		
\def\mnras{\jref@jnl{MNRAS}}		
\def\pra{\jref@jnl{Phys.~Rev.~A}}	
\def\prb{\jref@jnl{Phys.~Rev.~B}}	
\def\prc{\jref@jnl{Phys.~Rev.~C}}	
\def\prd{\jref@jnl{Phys.~Rev.~D}}	
\def\pre{\jref@jnl{Phys.~Rev.~E}}	
\def\prl{\jref@jnl{Phys.~Rev.~Lett.}}	
\def\pasp{\jref@jnl{PASP}}		
\def\pasj{\jref@jnl{PASJ}}		
\def\qjras{\jref@jnl{QJRAS}}		
\def\skytel{\jref@jnl{S\&T}}		
\def\solphys{\jref@jnl{Sol.~Phys.}}	
\def\sovast{\jref@jnl{Soviet~Ast.}}	
\def\ssr{\jref@jnl{Space~Sci.~Rev.}}	
\def\zap{\jref@jnl{ZAp}}			
\def\nat{\jref@jnl{Nature}}		
\def\iaucirc{\jref@jnl{IAU~Circ.}}
\def\aplett{\jref@jnl{Astrophys.~Lett.}}
\def\apspr{\jref@jnl{Astrophys.~Space~Phys.~Res.}}
\def\bain{\jref@jnl{Bull.~Astron.~Inst.~Netherlands}}
\def\fcp{\jref@jnl{Fund.~Cosmic~Phys.}}
\def\gca{\jref@jnl{Geochim.~Cosmochim.~Acta}}
\def\grl{\jref@jnl{Geophys.~Res.~Lett.}}
\def\jcp{\jref@jnl{J.~Chem.~Phys.}}	
\def\jgr{\jref@jnl{J.~Geophys.~Res.}}	
\def\jqsrt{\jref@jnl{J.~Quant.~Spec.~Radiat.~Transf.}}
\def\memsai{\jref@jnl{Mem.~Soc.~Astron.~Italiana}}
\def\nphysa{\jref@jnl{Nucl.~Phys.~A}}
\def\physrep{\jref@jnl{Phys.~Rep.}}
\def\physscr{\jref@jnl{Phys.~Scr}}
\def\planss{\jref@jnl{Planet.~Space~Sci.}}
\def\procspie{\jref@jnl{Proc.~SPIE}}

\def\ion#1#2{#1$\;${\small\rm\@Roman{#2}}\relax}

\newcount\lecurrentfam
\def\LaTeX{\lecurrentfam=\the\fam \leavevmode L\raise.42ex
\hbox{$\fam\lecurrentfam\scriptstyle\kern-.3em A$}\kern-.15em\TeX}
\def\sizrpt{
(\fontname\the\font): em=\the\fontdimen6\font, ex=\the\fontdimen5\font
\typeout{
(\fontname\the\font): em=\the\fontdimen6\font, ex=\the\fontdimen5\font
}}
\usepackage{sidecap}
\definecolor{dark}{rgb}{ 0.34, 0.09, 0.49}

\usepackage{fancyhdr,url}
\usepackage{transparent}
\usepackage{tablefootnote}
\usepackage{mathpazo}

\begin{document}

\noindent
\begin{raggedright}
{\Huge\bfseries Dust at cosmic dawn  \\[0.5em]
\Large Unveiling the onset of dust and obscured star formation at $z\gtrsim8$
}\end{raggedright}\\

\noindent Hiddo Algera$^{1}$, Tom Bakx$^{2}$, Manuel Aravena$^{3}$, Leindert Boogaard$^{4}$, Karina Caputi$^{5}$, Stefano Carniani$^{6}$, Pierre Cox$^{7}$, Elisabete da Cunha$^{8}$, Miroslava Dessauges-Zavadsky$^{9}$, Tanio Diaz Santos$^{10}$, Andreas Faisst$^{11}$, 
Andrea Ferrara$^{6}$,
Yoshinobu Fudamoto$^{12}$, Carlotta Gruppioni$^{13}$, Mahmoud Hamed$^{14}$, Jacqueline Hodge$^{4}$, Ryota Ikeda$^{4}$, Hanae Inami$^{15}$, Adarsh Kumar$^{16}$, Andrea Lapi$^{17}$, Daizhong Liu$^{18,19}$, Umberto Maio$^{20,21}$, Dieu Nguyen$^{22}$, Omima Osman$^{20,23}$, Andrea Pallottini$^{6,24}$, Gergö Popping$^{25}$, Francesco Salvestrini$^{20,21}$, Irene Shivaei$^{14}$, Laura Sommovigo$^{26,27}$, Justin Spilker$^{28}$, Livia Vallini$^{13}$, Julie Wardlow$^{29}$, Joris Witstok$^{30,31}$, Mengyuan Xiao$^{9}$

{\footnotesize{\vspace{1em}\noindent 
\textit{Affiliations can be found after the references}
\vspace{0.5em}
}}

\begin{tcolorbox}[boxsep=3pt,left=0pt,right=0pt,top=1pt,bottom=1pt,colframe=boxcolor,colback=boxcolor]
\justify
\textbf{Abstract:} Although less than 1 percent of the baryonic mass budget of a galaxy, cosmic dust profoundly alters the multi-wavelength appearance of galaxies across cosmic time. The (sub)millimeter wavelength regime is uniquely capable of weighing these dust reservoirs, and revealing the total extent of galaxy evolution that remains hidden behind them. With current facilities, however, doing so at $z > 8$ is largely out of reach. ALMA2040, a next-generation (sub)millimeter interferometer, is set to change this. Through an order-of-magnitude increase in sensitivity compared to ALMA, ALMA2040 is a transformational facility that will be able to routinely detect dust emission from galaxies at $z > 8$ and, through simultaneous multi-band characterization, accurately measure their dust masses, temperatures, and infrared luminosities. Combined with ancillary metallicity measurements -- including from the far-infrared lines detected by ALMA2040 itself -- this will establish the relative roles of supernovae and grain growth in building up dust in the earliest galaxies. Simultaneously, ALMA2040 will measure the fraction of star formation that is obscured by dust at cosmic dawn, and thereby the extent to which rest-UV and optical surveys miss early galaxy build-up. Without a 10$\times$ upgrade to ALMA -- comparable to the leap from the VLT to the ELT -- we will not be able to capitalize on the momentous advances enabled by \textit{JWST}, and will remain blind to the cold and obscured component of galaxy formation in the $z > 8$ Universe.
\end{tcolorbox}

\newpage{}

\vspace{-0.7cm}
\section{Scientific context and open questions} \vspace*{-0.3cm}

Dust fundamentally shapes the evolution of galaxies and their multi-wavelength appearance. The bulk of cosmic star formation is known to be dust-obscured at $z\lesssim4$ \citep[e.g.,][]{Casey2018,Zavala2021}, and its contribution remains non-negligible out to at least $z\sim7$ \citep[][]{Algera2023,Fujimoto2024,Traina2026}. With ALMA, individual dusty galaxies have now been detected as far back as $z=8.3$, a mere $\sim600\,$Myr after the Big Bang \citep[][]{Tamura2019,Bakx2020}, while observations with the \textit{JWST} indicate the presence of a dust-attenuated galaxy population even at $z > 10$ \citep[][]{Mitsuhashi2026b,Rodighiero2026}.

\begin{SCfigure}[5.3][htbp]
    \centering 
    \includegraphics[width=0.55\textwidth]{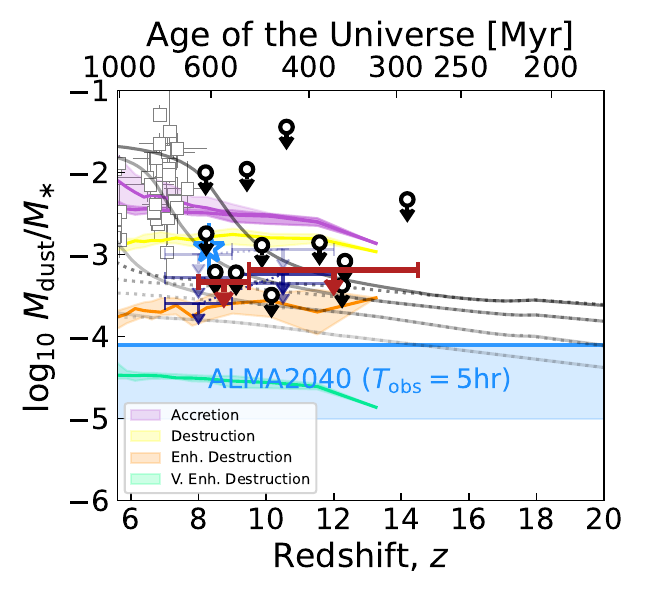}
    \hspace*{-0.35cm}\vspace*{-1.1cm}
    \caption{\textbf{The dust-to-stellar mass ratio as a function of cosmic time in the $z > 6$ Universe}, adapted from \citet{Bakx2026}. All individual galaxies at $z > 8$ remain undetected in dust continuum (black points; except Y1 shown as a blue star), and even in deep stacks (red and blue limits; \citealt{Ciesla2025}). These limits, as well as theoretical models \citep{Esmerian2024,Toyouchi2025}, suggest rapid and efficient dust growth in the $z < 8$ Universe, but are heavily affected by current sensitivity limits and selection effects. {\it ALMA2040 will provide the sensitivity increase needed to uncover the early mechanisms of dust build-up.}}
    \label{fig:MdustMstar} 
\end{SCfigure}

Understanding the origin and abundance of dust, as well as the importance of obscured star formation in the early Universe, requires pushing the detection and characterization of dusty galaxies deep into the epoch of \textit{cosmic dawn} (here defined as $z\gtrsim8$, effectively the current `dusty frontier'; Figure~\ref{fig:MdustMstar}). This requires a transformational (sub)millimeter facility capable of answering the following fundamental questions:

\begin{itemize}
    \vspace*{-0.2cm}
    \setlength{\itemsep}{-0.1cm}
    \item \textbf{When and how do the earliest dusty galaxies arise?}
    \item \textbf{What is the nature of early dust, and where in galaxies does it reside?}
    \item \textbf{What fraction of cosmic star formation remains obscured at cosmic dawn?}
    \vspace*{-0.2cm}
\end{itemize}

\noindent To achieve this, such a facility must satisfy two key requirements: an order-of-magnitude leap in raw sensitivity, and the ability to observe simultaneously across multiple bands, both of which are detailed in Section \ref{sec:technical_requirements}.

\vspace*{-0.3cm}\subsection{When and how do the earliest dusty galaxies arise?} \vspace*{-0.2cm}

\noindent How, and how rapidly, dust is produced in the early Universe remains a key open question. The first dust is thought to originate from core-collapse supernovae \citep[SNe;][]{TodiniFerrara2001}, though it remains unclear whether they alone account for the entirety of the earliest dust reservoirs. In the local Universe, dust production through Asymptotic Giant Branch (AGB) stars and growth in the interstellar medium (`grain growth') are believed to be the dominant channels \citep[e.g.,][]{DeLooze2020,Galliano2021}. However, at high redshift, AGB dust production is limited by both timescale and metallicity: more massive AGB stars ($\gtrsim3\,M_\odot$) evolve quickly but produce little dust at low metallicity, while lower-mass stars form dust largely independent of metallicity, but only after several hundred Myr \citep{Nanni2014,SchneiderMaiolino2024}. At $z\gtrsim8$, AGB stars are therefore expected to play a limited role in dust production.

ALMA observations of (UV-luminous and generally massive) galaxies at $6 \lesssim z \lesssim 8$ have revealed dust emission in a significant fraction of these systems \citep{Watson2015,Tamura2019,Inami2022}. Subsequent multi-band follow-up has enabled improved constraints on their dust masses \citep[e.g.,][]{Bakx2021}, which, combined with stellar mass measurements, provide a first-order diagnostic of their dust production pathways. The resulting dust-to-stellar mass ratios of $M_d / M_\star \sim 10^{-3} - 10^{-2}$ \citep[e.g.,][]{Witstok2022,Algera2024b} suggest either particularly high SN yields, an early onset of rapid AGB enrichment, or a significant contribution from grain growth \citep[e.g.,][]{Popping2017,DiCesare2023,Choban2025}. 

Disentangling the relative importance of these channels requires additional ancillary data beyond stellar mass alone. Metallicity, in particular, offers a key diagnostic \citep[e.g.,][]{RemyRuyer2014,DeVis2019,Geesink2026,Tailor2026}: while dust production via SNe alone is expected to yield a dust-to-gas (DTG) ratio that scales roughly linearly with metallicity (i.e., a constant dust-to-metal ratio, DTM), the onset of efficient grain growth is predicted to produce a much steeper, super-linear increase in DTG above a characteristic metallicity threshold of $Z\gtrsim0.1-0.2\,Z_\odot$ \citep[e.g.,][]{Asano2013,Choban2024}. With the advent of the \textit{JWST}, high-redshift constraints on the metallicity-dependence of the DTG and DTM ratios are now possible at $z \gtrsim 2.5$. While statistics of dusty galaxies with metallicity information remain limited, the first constraints suggest that metal-rich ($Z\gtrsim0.2\,Z_\odot$) galaxies in the Epoch of Reionization host substantial dust reservoirs but are located (slightly) below the local DTG-metallicity scaling relation \citep[][]{Heintz2025,Umehata2025,Algera2026a,Rowland2026_dla,Pollock2026}. This suggests that ISM grain growth, or possibly AGB star production, contributes to overall dust build-up at this epoch, but may not yet be fully efficient.

To probe what are likely `pristine' SN-produced dust reservoirs, it is essential to constrain the dust masses of ultra-high-redshift ($z\gtrsim10$) and/or metal-poor ($Z\lesssim 0.1\,Z_\odot$) galaxies, where AGB production and grain growth should both be inefficient. Beyond $z = 8$, however, MACS0416-Y1 at $z=8.31$ \citep{Tamura2019,Bakx2020} has remained the only known dusty galaxy since its discovery in 2019, even though \textit{JWST} indicates the presence of dust-attenuated galaxies at even earlier times \citep{Donnan2025,Mitsuhashi2026b}. Several recent deep ALMA-based searches at $z > 8$ have come up empty-handed \citep[][]{Algera2026b,Bakx2026,Mitsuhashi2026}, suggesting a rapid transition in the typical dust contents of galaxies between $z\sim8-9$, the origin of which remains heavily debated \citep[e.g.,][]{Ferrara2023,Burgarella2025,Narayanan2026,Sommovigo2026}. Current ALMA upper limits on individual galaxies suggest $M_d/M_\star \lesssim 10^{-3} - 10^{-2.5}$ (Figs.\ \ref{fig:MdustMstar} and \ref{fig:DTG_DTS}), though with the caveat that the exact limits depend sensitively on the assumed dust temperature. At face value, these imply supernova yields below $y_\mathrm{SN} \lesssim 0.1 - 0.3\,M_\odot\,\mathrm{SN}^{-1}$ given a standard Chabrier IMF and no further dust production/destruction mechanisms \citep{Michalowski2015}. However, to constrain SN yields and reveal the importance of grain growth at high redshift, much deeper observations across multiple bands are essential, probing down to $M_d / M_\star \lesssim 10^{-4}$ (equivalent to $y_\mathrm{SN} \lesssim 10^{-2}\,M_\odot\,\mathrm{SN}^{-1}$). As quantified below, ALMA2040 will deliver these constraints even in typical, low-mass galaxies ($M_\star \sim 10^{8.5-9}\,M_\odot$) at cosmic dawn.

\begin{figure}[t!]
    \centering
    \includegraphics[width=0.98\textwidth]{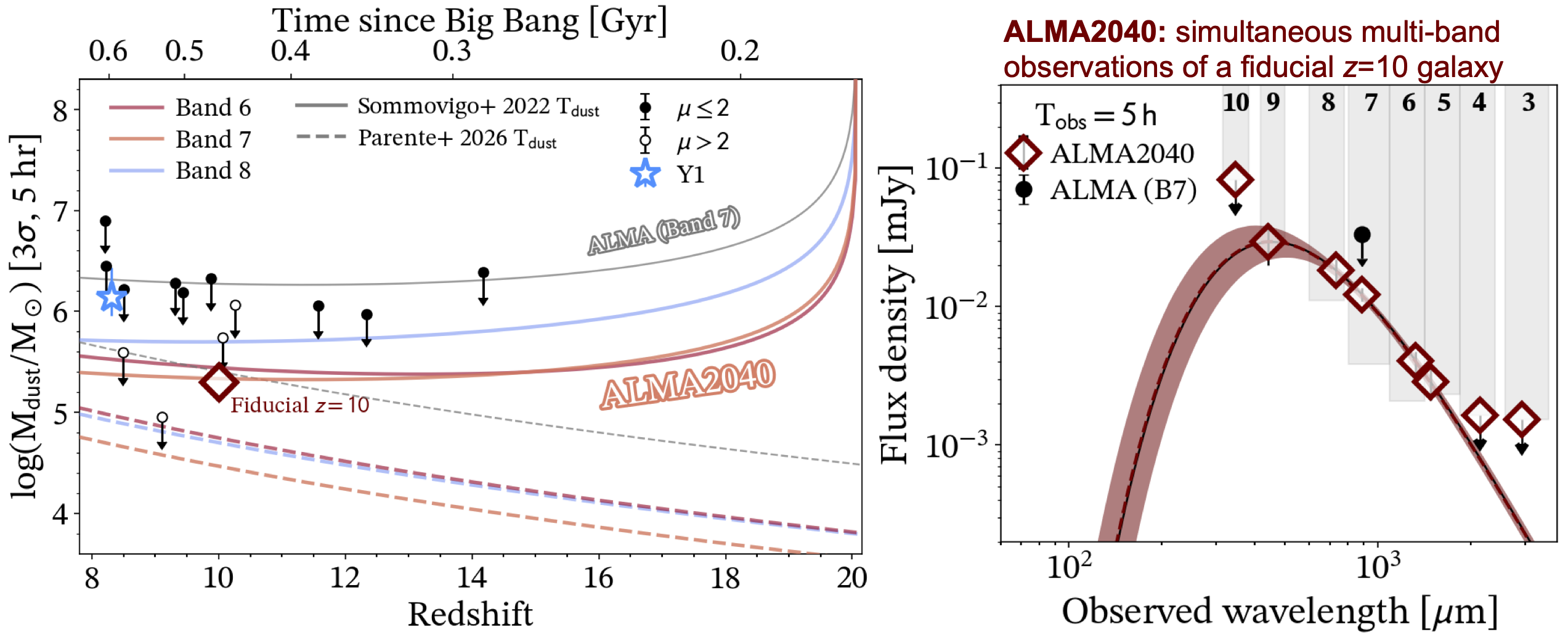}
    \caption{\textit{Left:} \textbf{Predicted dust mass sensitivity of ALMA2040 in Bands 6, 7 and 8 $(3\sigma, 5\,\mathrm{h})$}, for a single-temperature, optically thin modified blackbody with $\beta_\mathrm{IR}=2$. The limits depend strongly on the assumed dust temperature, so we show two $T_d - z$ relations that bracket a plausible range: \citet[$T_d(z=10)\sim50\,\mathrm{K}$; solid]{Sommovigo2022} and \citet[$T_d(z=10)\sim95\,\mathrm{K}$; dashed]{Parente2026}. ALMA Band 7 limits are shown in grey for comparison. Existing dust mass limits at $z>8$ and the sole detection at this epoch (Y1 at $z=8.31$) are split by lensing magnification ($\mu \lessgtr 2$). While ALMA has provided deep limits for a few rare, strongly lensed sources \citep[e.g., JD1 at $z=9.11$;][]{Hashimoto2019}, ALMA2040 will provide such constraints for normal, unlensed galaxies out to even earlier epochs. \textit{Right:} \textbf{constraints on the dust SED of a fiducial $\mathbf{z=10}$ galaxy $\mathbf{(M_d, T_d) = (2\times10^5\,M_\odot, 70\,\mathrm{K})}$}, marked by the maroon diamond in the left panel. Grey shading shows the $3\sigma$, $5\,\mathrm{h}$ detection limit in each band. A single ALMA band yields only an upper limit (black point), whereas ALMA2040 observing Bands $1$--$10$ simultaneously would detect the dust emission in five of them, directly constraining the dust mass, temperature and infrared luminosity through a MBB fit (maroon).}
    \label{fig:sensitivity}
\end{figure}

\vspace*{-0.1cm}\subsubsection{An order-of-magnitude leap in continuum sensitivity with ALMA2040}\vspace*{-0.2cm}
\label{sec:sensitivity}

The left panel of Figure \ref{fig:sensitivity} shows the order-of-magnitude increase in sensitivity that ALMA2040 will deliver compared to ALMA -- currently the most sensitive facility at these wavelengths. Our sensitivity estimates assume a single-temperature, optically thin modified blackbody (MBB) with a fixed opacity of $\kappa_\nu=30\,\mathrm{cm}^2\mathrm{g}^{-1}$ at $\lambda_\mathrm{rest}=100\,\mu\mathrm{m}$ and an emissivity index $\beta_\mathrm{IR}=2.0$. With these assumptions, the limiting dust mass depends primarily on the typical dust temperature of high-redshift galaxies, which is poorly constrained beyond $z\gtrsim4$ \citep[e.g.,][]{Schreiber2018,Viero2022}. Uncertainties in the dust opacity, optical depth, and multi-temperature dust are not included \citep[see e.g.,][]{DaCunha2021,SommovigoAlgera2025,Fanciullo2026}, although multi-band data near the peak of the dust SED will help constrain the latter two.

In general, dust temperatures are believed to increase towards earlier times, and given that hotter dust is more luminous ($L_\mathrm{IR} \propto M_d T_d^{4+\beta_\mathrm{IR}}$), this improves the detectability of a fixed dust mass at higher redshifts -- provided that $T_d$ well exceeds the CMB temperature, which we account for following \citet{DaCunha2013}. Under plausible {evolving} dust temperature scenarios \citep{Sommovigo2022,Parente2026}, ALMA2040 will reach limiting dust masses of $\log(M_d/M_\odot)\sim3.9-5.4$ across $z\sim8-15$ in just $5\,\mathrm{h}$ on-source in Band 7 -- an order of magnitude deeper than ALMA can achieve in the same time. In other words, reaching this depth with ALMA today would require $\sim100\times$ longer integrations, making a routine $5\,\mathrm{h}$ ALMA2040 observation equivalent to several current-day Large Programs.

Particularly transformational would be the ability to conduct simultaneous multi-band observations. The right panel of Figure \ref{fig:sensitivity} demonstrates that a $5\,\mathrm{h}$ ALMA2040 observation would detect the dust emission of a fiducial $z=10$ galaxy with $(M_d, T_d, L_\mathrm{IR}) = (2 \times 10^5\,M_\odot,\, 70\,\mathrm{K},\, 10^{10.9}\,L_\odot)$ in five bands at once, enabling direct constraints on its dust mass, temperature, emissivity index, and infrared luminosity; obtaining similar constraints with ALMA would require over $2000\,\mathrm{h}$. Our ALMA2040 estimates are moreover conservative, as they assume a bandwidth of just $8\,\mathrm{GHz}$, whereas multi-band observations would likely be paired with expanded bandwidths (Section \ref{sec:technical_requirements}).

\begin{SCfigure}[50][t!]
    \centering
    \includegraphics[width=0.62\textwidth]{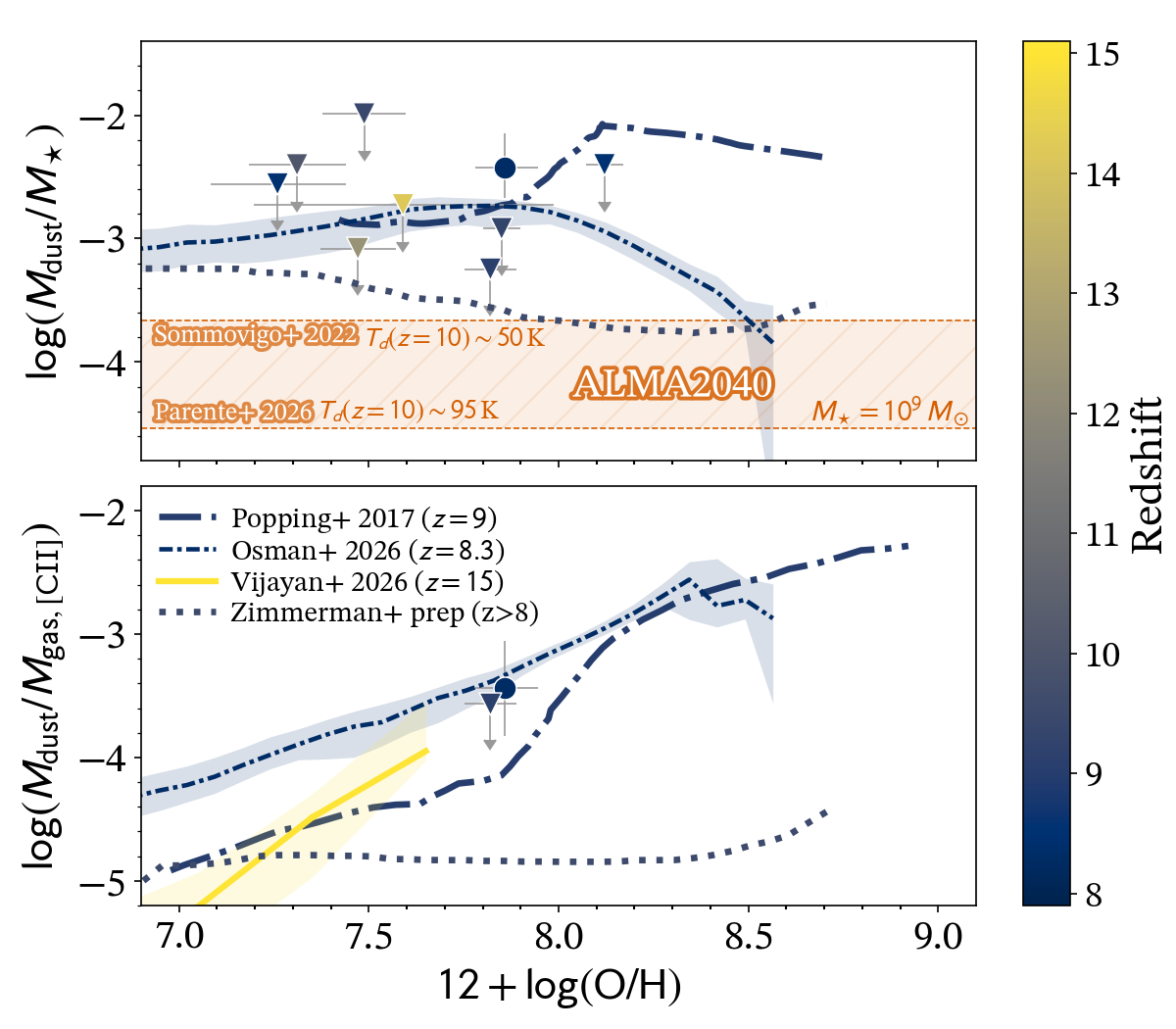}
    \caption{\textbf{The dust-to-stellar mass ratio (\textit{top}) and dust-to-gas ratio (\textit{bottom}) against metallicity.} The few existing constraints from $z > 8$ galaxies are compared to theoretical predictions \citep[][Zimmerman et al.\ in prep]{Popping2017,Osman2026,Vijayan2026}. The top panel shows ALMA2040's sensitivity in a $5\,\mathrm{h}$ Band 7 observation for a characteristic $M_\star = 10^{9}\,M_\odot$ and two $T_d - z$ relations (cf.\ Figure~\ref{fig:sensitivity}). Only with a $\sim10\times$ increase in sensitivity can the dust content of low-mass, low-metallicity galaxies at cosmic dawn be directly probed, distinguishing early dust formation pathways.} \label{fig:DTG_DTS}
\end{SCfigure}

\vspace*{-0.1cm}\subsubsection{Ultra-deep constraints on dust-to-gas and dust-to-stellar mass ratios at cosmic dawn}\vspace*{-0.2cm}

\noindent Detecting and characterizing low-mass dust reservoirs through multi-band data is crucial to understanding the origin of the earliest dust. Figure \ref{fig:DTG_DTS} shows the dust-to-stellar mass (DTS; top) and DTG ratios (bottom) as a function of metallicity, alongside predictions from the \textit{Santa Cruz} and \textit{GAEA} semi-analytical models \citep{Popping2017,Osman2026}, the \textit{Cosmic Sands} zoom-in simulations \citep[][Zimmerman et al.\ in prep.]{Narayanan2026}, and the \textit{COLIBRE} simulations \citep{Vijayan2026}. These models differ substantially in the efficiency and timing of dust production, destruction, and growth: for instance, grain growth is already efficient by $z\approx9$ in \textit{Santa Cruz} and \textit{GAEA}, but only sets in at $z\lesssim8$ in \textit{Cosmic Sands}, resulting in widely different DTS and DTG ratios. Telling these models apart requires measuring both ratios as a function of metallicity -- the key variable driving galaxy dust content -- at $z>8$, where hardly any constraints currently exist.

The DTS ratio, shown in the top panel, is a less direct proxy of dust build-up than the DTG ratio, but has the advantage of not requiring an ancillary gas mass measurement. At $z=10 - 15$, ALMA2040 will reach typical dust-to-stellar mass ratios of $M_d / M_\star \lesssim 10^{-4}$ for a $M_\star = 10^9\,M_\odot$ galaxy in just $5\,\mathrm{h}$ -- a stellar mass comparable to that of UV-bright galaxies such as GHZ2 \citep{Castellano2024,Zavala2024,Zavala2025} and JADES-GS-z14-0 \citep{Carniani2024,Carniani2025,Schouws2025}, where dust remains undetected to date at an order-of-magnitude shallower depth. Strongly gravitationally lensed galaxies at $z\gtrsim10$ \citep[e.g.,][]{Bradac2025} offer a complementary route to probing dust build-up in even fainter, lower-mass and typically metal-poor systems.

The DTG ratio, shown in the bottom panel, is a more direct probe of dust build-up, but requires ancillary gas masses. The observational data shown here use \cii{}-based gas masses following \citet{Zanella2018}; to date, however, the \cii{} line has only been confidently detected in a single $z > 8$ galaxy \citep{Bakx2020}, and tentatively in two others \citep[][]{Carniani2020,Fujimoto2024_ID4590}. The companion White Paper (WP) by \textit{Vallini et al.\ }shows how ALMA2040 will push \cii{} detections to significantly earlier cosmic times, providing insights into the cold gas contents of statistical samples of normal galaxies. By combining \cii{} with dynamical gas masses from high-resolution emission line kinematics \citep[e.g.,][]{Rowland2024,AlgeraHerreraCamus2025} and other tracers such as CO and \ci{} \citep[e.g.,][]{Cescon2026}, ALMA2040 will finally make it possible to populate this DTG-metallicity diagram at $z\gtrsim8$. Each of these gas tracers carries systematic uncertainties -- line-to-gas conversion factors depend on metallicity and ISM conditions, while dynamical masses rely on assumptions about the geometry and the dark matter contribution -- but measuring several of them in the same galaxies will allow these to be cross-checked. Through detailed measurements of the DTS and DTG ratios, ALMA2040 will discriminate between the plausible early dust-evolution pathways illustrated by the models in Figure \ref{fig:DTG_DTS}, across a wide range of stellar masses and metallicities, and provide the empirical benchmark against which future theoretical models of early dust production must be tested.

\vspace*{-0.1cm}\subsection{The nature and spatial distribution of early dust reservoirs}\vspace*{-0.2cm}

The apparent rapid transition in galaxy dust content between $z\sim8 - 9$ could reflect a change in the dominant production channel, i.e., SNe vs.\ grain growth. Beyond differing yields and efficiencies, these two mechanisms likely also produce dust with distinct physical properties. SNe have been suggested to primarily produce large grains, which attenuate UV and optical light with comparable efficiency, producing flat extinction curves rather than the steep curves characteristic of smaller grains \citep{HirashitaAoyama2019,Otaki2026,Shivaei2026,Sommovigo2026}. Grain growth, on the other hand, is believed to be most efficient on smaller dust grains, which may themselves be produced by shattering of larger grains in the ISM \citep[e.g.,][]{Hirashita2012}. Such interplay between shattering and growth, in addition to effects related to the star-dust geometry, is likely required to produce the typical attenuation curves seen in both the nearby and high-redshift Universe \citep[e.g.,][]{Vijayan2024,Matsumoto2025,Parente2026_grainsizes}.

Because the grain size distribution shapes how dust absorbs and subsequently re-radiates starlight, constraining the attenuation curve and dust continuum emission jointly -- for example leveraging energy-balance techniques -- offers a direct handle on grain properties, and by extension their production pathways. The combination of ALMA2040's sensitivity with ancillary rest-UV and -optical measurements will enable precisely this type of constraint in the first galaxies, complementary to current \textit{JWST}-based studies relying on rest-UV/optical spectra alone \citep[e.g.,][]{Fisher2025,Markov2025}. Pushing joint ALMA+\textit{JWST} (or ELT; Section \ref{sec:synergies}) studies to spatially resolved scales will moreover allow studying the impact of geometrical effects on the shape of the attenuation curve.

ALMA2040 will also enable detailed studies of the dust emissivity index, $\beta_\mathrm{IR}$, which provides additional insights into the size distribution and chemical composition of high-redshift dust. While the direct link between $\beta_\mathrm{IR}$ and dust properties is complex, it has been suggested that steep values of $\beta_\mathrm{IR} \gtrsim 2$ trace cold, dense regions where grain growth is likely to occur \citep[e.g.,][]{Kohler2012,Algera2024b}. While measurements of $\beta_\mathrm{IR}$ are currently limited to only the brightest galaxies and quasars at $z\gtrsim6$ \citep[e.g.,][]{Witstok2023,Tripodi2024}, ALMA2040 -- with its simultaneous multi-band capabilities -- will push these studies to more typical and more distant galaxies. Any evidence for evolution in the average $\beta_\mathrm{IR}$ with redshift would then be a likely signpost of evolution in galaxy dust properties.

It has also been suggested that the seemingly abrupt drop in galaxy dust content beyond $z\gtrsim8$ is due to dust being efficiently removed from highly starbursting galaxies through outflows \citep[][]{Ferrara2023}. The clearest evidence for such outflows will come from kinematics, such as through broad emission line wings in \cii{} and \oiiil{} \citep[e.g.,][see also \textit{Vallini et al.}]{Ginolfi2020,Akins2022}. In addition, for sufficiently luminous galaxies at cosmic dawn, ALMA2040 may make it possible to detect cold (turbulent) outflows in absorption against their dust continuum emission through tracers such as $\mathrm{OH}$, $\mathrm{OH}^+$ and $\mathrm{CH}^+$ (WP by \textit{Butler et al.}). Where such outflows are co-spatial with extended or offset dust continuum emission, this would offer direct evidence that dust itself -- not just the ionized or molecular gas -- is being ejected from these galaxies.

Testing this scenario further requires spatially resolved dust continuum observations coupled with sufficient sensitivity to low-surface-brightness emission -- a combination that remains challenging with current facilities. While ALMA has been able to observe a handful of galaxies in the Epoch of Reionization at $\lesssim 0.2''$ resolution \citep[e.g.,][]{Tamura2023,Rowland2024}, such observations are currently limited to a small number of exceptionally bright systems. Only with an order-of-magnitude increase in sensitivity can the dust emission of `normal' high-redshift galaxies be mapped on sub-kpc scales. Beyond outflows, the detection of spatial offsets between dust and stellar emission may also be a signpost of deeply embedded star formation missed in rest-UV and -optical observations \citep[e.g.,][]{Ocvirk2025,Rowland2026}.

\vspace*{-0.1cm}\subsection{What fraction of the high-redshift SFRD is dust-obscured?} \vspace*{-0.2cm}

\begin{figure}[t!]
    \centering
    \vspace{0.1cm}
    \includegraphics[width=0.95\textwidth]{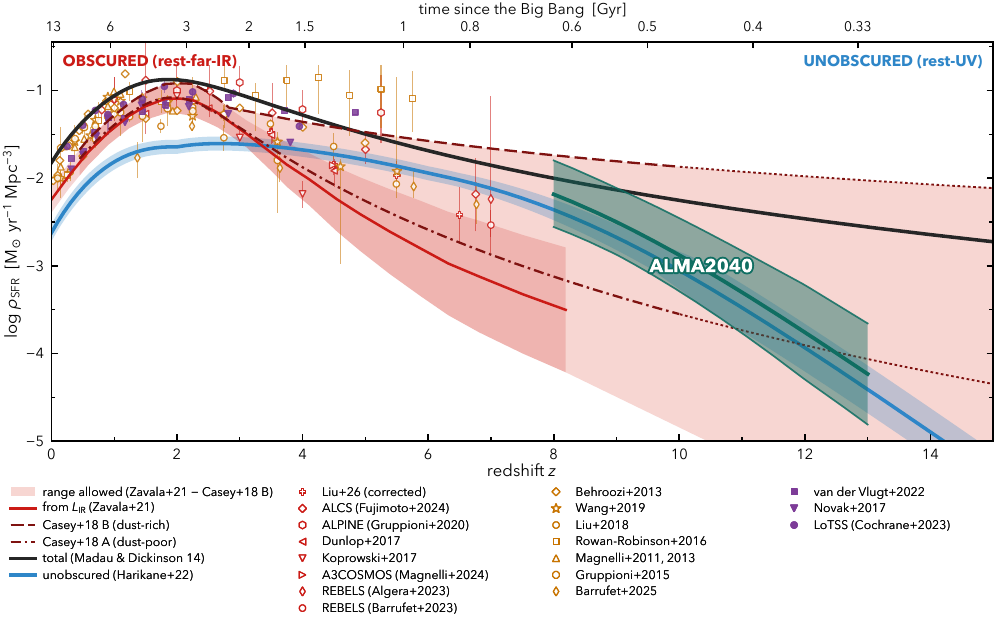}
    \caption{\textbf{The obscured cosmic star-formation rate density is unconstrained at high redshift.} Obscured (red) and unobscured (blue) cosmic star-formation rate density, with individual measurements from the compilation by \citet{Liu2026}. The obscured component is directly measured only out to $z\sim6$; beyond this, existing studies allow anything between the lower error edge of the blank-field dusty galaxy census of \citet{Zavala2021} and the dust-rich scenario of \citet{Casey2018}. The unobscured component, on the other hand, is constrained by \textit{JWST} to a few tenths of a dex. {\it ALMA2040 will be able to detect $L_*$ galaxies individually down to $f_{\rm obsc} = 0.1$, and thereby constrain the obscured contribution to the SFRD from the earliest galaxies.}}
    \label{fig:csfrd}
\end{figure}

Independent of its physical origin, dust hides a fraction of cosmic star formation from rest-UV and -optical view, $f_\mathrm{obsc} = \rho_\mathrm{SFR,obsc} / \rho_\mathrm{SFR,total}$ (Figure \ref{fig:csfrd}). The unobscured branch is now measured to high precision by \textit{JWST} \citep[e.g.,][]{Weibel2026}, but the obscured component is only directly measured via the infrared luminosity function out to $z\sim6$ \citep[e.g.,][]{Zavala2021}. Beyond this, indirect routes (e.g., \textit{JWST}-based samples; see \citealt{Liu2026} for a compilation) are subject to selection biases and are furthermore limited by systematic uncertainties on obscured SFRs due to the poorly constrained dust temperature (Section \ref{sec:sensitivity}). Such studies currently allow anything from a few percent to three quarters of cosmic star formation to be obscured by $z=10$. We therefore do not know when dust began to obscure our view of early galaxy build-up.

This question has gained additional urgency with the discovery of an unexpectedly abundant population of UV-luminous galaxies at $z>10$ \citep[e.g.,][]{Harikane2023,Castellano2024}. Proposed explanations include enhanced star-formation efficiencies, bursty star formation, top-heavy initial mass functions, and the efficient removal of dust from these systems \citep[e.g.,][]{Dekel2023,Ferrara2023}. Crucially, the UV luminosities underlying this tension are \textit{observed} quantities: any dust attenuation would exacerbate the tension with current galaxy formation models by implying even higher intrinsic luminosities and star-formation rates. ALMA2040 will reach SFR$_{\rm obsc}\approx1-7\,M_\odot\,\mathrm{yr}^{-1}$ at $z\approx10$ in a $5\,\mathrm{h}$ observation, sufficient to detect even modest obscured fractions in the UV-bright population, and thereby determine if these galaxies are truly dust-free, or how much higher the intrinsic UV luminosities that models must explain really are.

Because simultaneous multi-band photometry measures $T_d$ and $L_\mathrm{IR}$ for individual galaxies, ALMA2040 can constrain $\rho_\mathrm{SFR,obsc}(z)$ without assuming a single (evolving) dust temperature. To quantify this, we take the UV luminosity function from \citet{Harikane2023} and assume, ad hoc, that every galaxy is $60\%$ obscured. In a $5\,$h follow-up of UV-selected sources, ALMA2040 would then detect the dust emission of galaxies brighter than $M_\mathrm{UV} < -18.4$ at $z=9$ ($-18.8$ at $z=13$), constraining the obscured SFRD to within $0.4\,$dex at $z=9$ ($0.6\,$dex at $z=13$). Combined in quadrature with the uncertainty on the UV SFRD, this yields the forecast total SFRD in Figure~\ref{fig:csfrd}.

In parallel, blind studies -- such as scaled up versions of the ALMA Large Programs ASPECS \citep{Walter2016} and CHAMPS \citep{Faisst2026} -- will be able to find the more obscured sources missed from the UV-selected census (see \textit{Sun et al.}). Their abundance at cosmic dawn directly affects not only the SFRD, but also the high-mass end of the stellar mass function, which is currently derived exclusively from rest-UV/optically selected samples (e.g., \citealt{Weibel2024}). Because mapping speed scales with the square of the continuum sensitivity, ALMA2040 will survey areas $\sim100\times$ faster than ALMA to a given depth, enabling deep blind surveys out to $z\sim10$. Simultaneously, its wide instantaneous bandwidth allows the same observations to serve as spectral scans, enabling robust redshift confirmation of continuum-detected sources through \ciil{}, \oiiil{} and other far-infrared lines. Such surveys would discriminate between the dust-rich and dust-poor tracks of Figure~\ref{fig:csfrd}, which presently remain unconstrained.

\vspace*{-0.20cm}\section{ALMA2040: Technical Requirements} \vspace*{-0.3cm}
\label{sec:technical_requirements}

To study dust at the earliest epochs, ALMA2040 must deliver two transformational capabilities. The first is an order-of-magnitude leap in raw sensitivity, most readily attained through a large ($\sim10\times$) increase in collecting area relative to ALMA, combined with improved receiver technology and expanded continuum bandwidths. This is essential to detect dust emission in the typical low-mass, metal-poor galaxy population at cosmic dawn that currently lies beyond reach. This sensitivity moreover allows resolving dust emission on sub-kpc spatial scales, to map directly where it is being produced. While longer baselines are not strictly needed -- it is sensitivity rather than angular resolution that currently limits spatially resolved studies -- they would provide the possibility of studying dust on scales down to $\lesssim100\,\mathrm{pc}$ in the brightest $z > 8$ galaxies.

The second requirement for ALMA2040 is the ability to conduct simultaneous multi-band observations. As shown in Figure \ref{fig:sensitivity}, this would deliver a further leap in survey efficiency by removing the need for slow, often-secondary multi-band follow-up to reliably constrain dust masses and luminosities while simultaneously enabling a direct assessment of selection biases arising from source-to-source variations in dust temperature. Observing the bands together, through the same atmosphere, further reduces the band-to-band calibration error, and can reduce overheads. The ability to perform phase transfer, not just in time (\citealt{Maud2023}) but instead across frequency, will enable short-wavelength capabilities across more weather conditions.

\vspace*{-0.20cm}\section{Synergies with current and future facilities}
\label{sec:synergies} \vspace*{-0.3cm}

Studies of the earliest dusty galaxies with a next-generation (sub)millimeter telescope will benefit from powerful synergies with current and upcoming facilities. First of all, \textit{JWST}-selected samples of high-redshift galaxies will remain essential even once the telescope itself is retired, providing large samples of normal galaxies for ALMA2040 to follow up -- moving beyond the tip of the iceberg that is now being characterized by ALMA. Wide-area near-infrared surveys such as those by \textit{Euclid} \citep{Euclid_Allen2026} and \textit{Roman} will provide UV-luminous targets -- including strongly lensed systems -- out to $z\sim10$, while the GREX-PLUS space mission \citep{Inoue2022} will extend this selection to even earlier epochs ($z\gtrsim10-15$). ALMA2040's wide instantaneous bandwidth will be valuable for spectroscopically confirming high-redshift candidates through far-infrared lines, while simultaneously constraining their dust content.

Interpreting high-redshift dust detections will require further ancillary data. \textit{JWST} observations will remain crucial for the rest-UV and -optical characterization of high-redshift galaxies, delivering key quantities such as stellar mass, unobscured SFRs, and -- for a subset -- metallicities from ionized gas tracers. The ELT will complement this: together with \textit{JWST} and ALMA2040 itself, it will provide the (rest-UV) metallicity measurements needed to fully exploit gas mass and dust constraints in the context of early dust build-up, and the combination of ALMA2040 and the ELT will moreover map both dust emission and UV attenuation curves on resolved spatial scales. At longer wavelengths, radio facilities such as the Square Kilometre Array (SKA) and next-generation Very Large Array (ngVLA) will detect molecular gas in Reionization-era galaxies through (low-$J$) CO emission \citep[e.g.,][]{Wang2026_ska}, enabling independent constraints on galaxy gas masses and, in combination with ALMA2040, the dust-to-gas ratio.

Finally, wide-area, blind surveys with the \textit{PRIMA} space mission \citep{Glenn2025}, large single-dish (sub)millimeter facilities such as LMT/TolTEC and AtLAST (\citealt{vanKampen2024}), and radio telescopes such as the SKA \citep{Algera2026_ska,An2026_ska,Prandoni2026_ska} will complement ALMA2040 by constraining the SFRD over significantly larger areas at $z \lesssim 8$ .

\vspace*{-0.30cm}\section*{Acknowledgements} \vspace*{-0.30cm}
This work was developed as part of the broader ALMA2040 community effort. We acknowledge the Lorentz Center and NOVA for supporting the community workshops and coordination activities that helped initiate and shape the ALMA2040 White Paper process. We also thank the internal and external referees for their valuable comments that helped improve this White Paper. \\

\bibliographystyle{apj_mod}
\bibliography{bibliography}

\begin{thebibliography}{102}
\expandafter\ifx\csname natexlab\endcsname\relax\def\natexlab#1{#1}\fi

\bibitem[{{Akins} {et~al.}(2022){Akins}, {Fujimoto}, {Finlator}, {Watson}, {Knudsen}, {Richard}, {Bakx}, {Hashimoto}, {Inoue}, {Matsuo}, {Micha{\l}owski}, \& {Tamura}}]{Akins2022}
\textbf{Akins} {et~al.} 2022, \apj, 934, 64.
\bibitem[{{Algera} {et~al.}(2026{\natexlab{a}}){Algera}, {Weaver}, {Bakx}, {Aravena}, {Bouwens}, {Cescon}, {Chen}, {da Cunha}, {Dayal}, {Faisst}, {Ferrara}, {Fujimoto}, {Hashimoto}, {Heintz}, {Herrera-Camus}, {Hodge}, {Inami}, {Inoue}, {Matthee}, {Meyer}, {Mizukoshi}, {Mondal}, {Nanayakkara}, {Oesch}, {Pallottini}, {R{\"o}ttgering}, {Rowland}, {Schouws}, {Smit}, {Sommovigo}, {Stark}, {Sugahara}, {Vallini}, {Vijarnwannaluk}, {van der Werf}, {Werner}, {Witstok}, \& {Xiao}}]{Algera2026b}
\textbf{Algera} {et~al.} 2026{\natexlab{a}}, The Open Journal of Astrophysics, 9, 65333.
\bibitem[{{Algera} {et~al.}(2025){Algera}, {Herrera-Camus}, {Aravena}, {Assef}, {Bakx}, {Bolatto}, {Cescon}, {Chen}, {da Cunha}, {Dayal}, {De Looze}, {Diaz-Santos}, {Faisst}, {Ferrara}, {F{\"o}rster Schreiber}, {Hathi}, {Ikeda}, {Inami}, {Jones}, {Koekemoer}, {Lutz}, {Rela{\~n}o}, {Romano}, {Rowland}, {Sommovigo}, {Vallini}, {Vijayan}, {Villanueva}, \& {van der Werf}}]{AlgeraHerreraCamus2025}
\textbf{Algera} {et~al.} 2025, arXiv e-prints, arXiv:2512.02320.
\bibitem[{{Algera} {et~al.}(2024){Algera}, {Inami}, {De Looze}, {Ferrara}, {Hirashita}, {Aravena}, {Bakx}, {Bouwens}, {Bowler}, {Da Cunha}, {Dayal}, {Fudamoto}, {Hodge}, {Hygate}, {van Leeuwen}, {Nanayakkara}, {Palla}, {Pallottini}, {Rowland}, {Smit}, {Sommovigo}, {Stefanon}, {Vijayan}, \& {van der Werf}}]{Algera2024b}
\textbf{Algera} {et~al.} 2024, \mnras, 533, 3098.
\bibitem[{{Algera} {et~al.}(2023){Algera}, {Inami}, {Oesch}, {Sommovigo}, {Bouwens}, {Topping}, {Schouws}, {Stefanon}, {Stark}, {Aravena}, {Barrufet}, {da Cunha}, {Dayal}, {Endsley}, {Ferrara}, {Fudamoto}, {Gonzalez}, {Graziani}, {Hodge}, {Hygate}, {de Looze}, {Nanayakkara}, {Schneider}, \& {van der Werf}}]{Algera2023}
\textbf{Algera} {et~al.} 2023, \mnras, 518, 6142.
\bibitem[{{Algera} {et~al.}(2026{\natexlab{b}}){Algera}, {Rowland}, {Stefanon}, {Palla}, {Sommovigo}, {Inami}, {Bouwens}, {Aravena}, {Bowler}, {Dayal}, {De Looze}, {Ferrara}, {Fisher}, {Graziani}, {Gulis}, {Heintz}, {Hodge}, {Laza-Ramos}, {van Leeuwen}, {Pallottini}, {Phillips}, {Schouws}, {Smit}, {Stark}, \& {van der Werf}}]{Algera2026a}
\textbf{Algera} {et~al.} 2026{\natexlab{b}}, \mnras, 545, staf1897.
\bibitem[{{Algera} {et~al.}(2026{\natexlab{c}}){Algera}, {Sargent}, {Murphy}, \& {An}}]{Algera2026_ska}
\textbf{Algera} {et~al.} 2026{\natexlab{c}}, Advancing Astrophysics with the SKA I, 71.
\bibitem[{{An} {et~al.}(2026){An}, {Tabatabaei}, {Seymour}, {Khademi}, {Yuan}, {Wang}, {Grundy}, {Sargent}, {Wang}, {Zhang}, {Zhao}, {Algera}, {Brinks}, \& {Vaccari}}]{An2026_ska}
\textbf{An} {et~al.} 2026, Advancing Astrophysics with the SKA I, 70.
\bibitem[{{Asano} {et~al.}(2013){Asano}, {Takeuchi}, {Hirashita}, \& {Inoue}}]{Asano2013}
\textbf{Asano} {et~al.} 2013, Earth, Planets and Space, 65, 213.
\bibitem[{{Bakx} {et~al.}(2026){Bakx}, {Algera}, {Jolly}, {Esmerian}, {Knudsen}, {Sommovigo}, {Witstok}, {Carniani}, {Chen}, {Eales}, {Ferrara}, {Fudamoto}, {Hagimoto}, {Hashimoto}, {Inami}, {Inoue}, {Khouri}, {Mitsuhashi}, {Nyman}, {Olander}, {Serjeant}, {Smit}, {Yoon}, {Zavala}, {Aalto}, {Casey}, {Tamura}, \& {Vlemmings}}]{Bakx2026}
\textbf{Bakx} {et~al.} 2026, \mnras, 546, staf2284.
\bibitem[{{Bakx} {et~al.}(2021){Bakx}, {Sommovigo}, {Carniani}, {Ferrara}, {Akins}, {Fujimoto}, {Hagimoto}, {Knudsen}, {Pallottini}, {Tamura}, \& {Watson}}]{Bakx2021}
\textbf{Bakx} {et~al.} 2021, \mnras, 508, L58.
\bibitem[{{Bakx} {et~al.}(2020){Bakx}, {Tamura}, {Hashimoto}, {Inoue}, {Lee}, {Mawatari}, {Ota}, {Umehata}, {Zackrisson}, {Hatsukade}, {Kohno}, {Matsuda}, {Matsuo}, {Okamoto}, {Shibuya}, {Shimizu}, {Taniguchi}, \& {Yoshida}}]{Bakx2020}
\textbf{Bakx} {et~al.} 2020, \mnras, 493, 4294.
\bibitem[{{Brada{\v{c}}} {et~al.}(2025){Brada{\v{c}}}, {Jude{\v{z}}}, {Willott}, {Rihtar{\v{s}}ic}, {Martis}, {Harshan}, {Felicioni}, {Asada}, {Desprez}, {Clowe}, {Gonzalez}, {Jones}, {Lemaux}, {Markevitch}, {Markov}, {Mowla}, {Noirot}, {Peter}, {Robertson}, {Sarrouh}, {Sawicki}, {Schrabback}, \& {Tripodi}}]{Bradac2025}
\textbf{Brada{\v{c}}} {et~al.} 2025, \apjl, 995, L74.
\bibitem[{{Burgarella} {et~al.}(2025){Burgarella}, {Buat}, {Theul{\'e}}, {Zavala}, {Dickinson}, {Arrabal Haro}, {Bagley}, {Boquien}, {Cleri}, {Dewachter}, {Ferguson}, {Fern{\`a}ndez}, {Finkelstein}, {Gawiser}, {Grazian}, {Grogin}, {Holwerda}, {Kartaltepe}, {Kewley}, {Kirkpatrick}, {Kocevski}, {Koekemoer}, {Long}, {Lotz}, {Lucas}, {Mobasher}, {Papovich}, {P{\'e}rez-Gonz{\`a}lez}, {Pirzkal}, {Ravindranath}, {Rodighiero}, {Roehlly}, {Rose}, {Seill{\'e}}, {Somerville}, {Wilkins}, {Yang}, \& {Yung}}]{Burgarella2025}
\textbf{Burgarella} {et~al.} 2025, \aap, 699, A336.
\bibitem[{{Carniani} {et~al.}(2025){Carniani}, {D'Eugenio}, {Ji}, {Parlanti}, {Scholtz}, {Sun}, {Venturi}, {Bakx}, {Curti}, {Maiolino}, {Tacchella}, {Zavala}, {Hainline}, {Witstok}, {Johnson}, {Alberts}, {Bunker}, {Charlot}, {Eisenstein}, {Helton}, {Jakobsen}, {Kumari}, {Robertson}, {Saxena}, {{\"U}bler}, {Williams}, {Willmer}, \& {Willott}}]{Carniani2025}
\textbf{Carniani} {et~al.} 2025, \aap, 696, A87.
\bibitem[{{Carniani} {et~al.}(2020){Carniani}, {Ferrara}, {Maiolino}, {Castellano}, {Gallerani}, {Fontana}, {Kohandel}, {Lupi}, {Pallottini}, {Pentericci}, {Vallini}, \& {Vanzella}}]{Carniani2020}
\textbf{Carniani} {et~al.} 2020, \mnras, 499, 5136.
\bibitem[{{Carniani} {et~al.}(2024){Carniani}, {Hainline}, {D'Eugenio}, {Eisenstein}, {Jakobsen}, {Witstok}, {Johnson}, {Chevallard}, {Maiolino}, {Helton}, {Willott}, {Robertson}, {Alberts}, {Arribas}, {Baker}, {Bhatawdekar}, {Boyett}, {Bunker}, {Cameron}, {Cargile}, {Charlot}, {Curti}, {Curtis-Lake}, {Egami}, {Giardino}, {Isaak}, {Ji}, {Jones}, {Kumari}, {Maseda}, {Parlanti}, {P{\'e}rez-Gonz{\'a}lez}, {Rawle}, {Rieke}, {Rieke}, {Del Pino}, {Saxena}, {Scholtz}, {Smit}, {Sun}, {Tacchella}, {{\"U}bler}, {Venturi}, {Williams}, \& {Willmer}}]{Carniani2024}
\textbf{Carniani} {et~al.} 2024, \nat, 633, 318.
\bibitem[{{Casey} {et~al.}(2018){Casey}, {Zavala}, {Spilker}, {da Cunha}, {Hodge}, {Hung}, {Staguhn}, {Finkelstein}, \& {Drew}}]{Casey2018}
\textbf{Casey} {et~al.} 2018, \apj, 862, 77.
\bibitem[{{Castellano} {et~al.}(2024){Castellano}, {Napolitano}, {Fontana}, {Roberts-Borsani}, {Treu}, {Vanzella}, {Zavala}, {Arrabal Haro}, {Calabr{\`o}}, {Llerena}, {Mascia}, {Merlin}, {Paris}, {Pentericci}, {Santini}, {Bakx}, {Bergamini}, {Cupani}, {Dickinson}, {Filippenko}, {Glazebrook}, {Grillo}, {Kelly}, {Malkan}, {Mason}, {Morishita}, {Nanayakkara}, {Rosati}, {Sani}, {Wang}, \& {Yoon}}]{Castellano2024}
\textbf{Castellano} {et~al.} 2024, \apj, 972, 143.
\bibitem[{{Cescon} {et~al.}(2026){Cescon}, {Hodge}, {Boogaard}, {Algera}, {Rowland}, {Riechers}, {Smit}, {De Looze}, {Bouwens}, {van der Werf}, {Aravena}, {da Cunha}, {Dayal}, {Ferrara}, {Fisher}, {Inami}, {Mancera Pi{\~n}a}, {Oesch}, {Pallottini}, {Rybak}, {Schouws}, {Sommovigo}, {Stefanon}, \& {Vallini}}]{Cescon2026}
\textbf{Cescon} {et~al.} 2026, \mnras, 549, stag924.
\bibitem[{{Choban} {et~al.}(2024){Choban}, {Kere{\v{s}}}, {Sandstrom}, {Hopkins}, {Hayward}, \& {Faucher-Gigu{\`e}re}}]{Choban2024}
\textbf{Choban} {et~al.} 2024, \mnras, 529, 2356.
\bibitem[{{Choban} {et~al.}(2025){Choban}, {Salim}, {Kere{\v{s}}}, {Hayward}, \& {Sandstrom}}]{Choban2025}
\textbf{Choban} {et~al.} 2025, \mnras, 537, 1518.
\bibitem[{{Ciesla} {et~al.}(2025){Ciesla}, {Adscheid}, {Magnelli}, {Boquien}, {Laporte}, {B{\'e}thermin}, {Carvajal}, {Schinnerer}, \& {Liu}}]{Ciesla2025}
\textbf{Ciesla} {et~al.} 2025, \aap, 700, A277.
\bibitem[{{da Cunha} {et~al.}(2013){da Cunha}, {Groves}, {Walter}, {Decarli}, {Weiss}, {Bertoldi}, {Carilli}, {Daddi}, {Elbaz}, {Ivison}, {Maiolino}, {Riechers}, {Rix}, {Sargent}, \& {Smail}}]{DaCunha2013}
\textbf{da Cunha} {et~al.} 2013, \apj, 766, 13.
\bibitem[{{da Cunha} {et~al.}(2021){da Cunha}, {Hodge}, {Casey}, {Algera}, {Kaasinen}, {Smail}, {Walter}, {Brandt}, {Dannerbauer}, {Decarli}, {Groves}, {Knudsen}, {Swinbank}, {Weiss}, {van der Werf}, \& {Zavala}}]{DaCunha2021}
\textbf{da Cunha} {et~al.} 2021, \apj, 919, 30.
\bibitem[{{De Looze} {et~al.}(2020){De Looze}, {Lamperti}, {Saintonge}, {Rela{\~n}o}, {Smith}, {Clark}, {Wilson}, {Decleir}, {Jones}, {Kennicutt}, {Accurso}, {Brinks}, {Bureau}, {Cigan}, {Clements}, {De Vis}, {Fanciullo}, {Gao}, {Gear}, {Ho}, {Hwang}, {Micha{\l}owski}, {Lee}, {Li}, {Lin}, {Liu}, {Lomaeva}, {Pan}, {Sargent}, {Williams}, {Xiao}, \& {Zhu}}]{DeLooze2020}
\textbf{De Looze} {et~al.} 2020, \mnras, 496, 3668.
\bibitem[{{De Vis} {et~al.}(2019){De Vis}, {Jones}, {Viaene}, {Casasola}, {Clark}, {Baes}, {Bianchi}, {Cassara}, {Davies}, {De Looze}, {Galametz}, {Galliano}, {Lianou}, {Madden}, {Manilla-Robles}, {Mosenkov}, {Nersesian}, {Roychowdhury}, {Xilouris}, \& {Ysard}}]{DeVis2019}
\textbf{De Vis} {et~al.} 2019, \aap, 623, A5.
\bibitem[{{Dekel} {et~al.}(2023){Dekel}, {Sarkar}, {Birnboim}, {Mandelker}, \& {Li}}]{Dekel2023}
\textbf{Dekel} {et~al.} 2023, \mnras, 523, 3201.
\bibitem[{{Di Cesare} {et~al.}(2023){Di Cesare}, {Graziani}, {Schneider}, {Ginolfi}, {Venditti}, {Santini}, \& {Hunt}}]{DiCesare2023}
\textbf{Di Cesare} {et~al.} 2023, \mnras, 519, 4632.
\bibitem[{{Donnan} {et~al.}(2025){Donnan}, {Dickinson}, {Taylor}, {Arrabal Haro}, {Finkelstein}, {Stanton}, {Jung}, {Papovich}, {Akins}, {Koekemoer}, {McLeod}, {Napolitano}, {Amor{\'\i}n}, {Begley}, {Burgarella}, {Carnall}, {Casey}, {Calabr{\`o}}, {Cullen}, {Dunlop}, {Ellis}, {Fern{\'a}ndez}, {Giavalisco}, {Hirschmann}, {Hu}, {Illingworth}, {Kartaltepe}, {Kocevski}, {Kokorev}, {Leung}, {Lucas}, {Morales}, {McLure}, {Pentericci}, {P{\'e}rez-Gonz{\'a}lez}, {Somerville}, {Stevenson}, {Trump}, {Yung}, \& {Zavala}}]{Donnan2025}
\textbf{Donnan} {et~al.} 2025, \apj, 993, 224.
\bibitem[{{Esmerian} \& {Gnedin}(2024)}]{Esmerian2024}
\textbf{Esmerian} {et~al.} 2024, \apj, 968, 113.
\bibitem[{{Euclid Collaboration} {et~al.}(2026){Euclid Collaboration}, {Allen}, {Oesch}, {Bowler}, {Toft}, {Matharu}, {Weaver}, {McPartland}, {Shuntov}, {Sanders}, {Mobasher}, {McCracken}, {Atek}, {Ba{\~n}ados}, {Barrow}, {Belladitta}, {Carollo}, {Castellano}, {Conselice}, {Eisenhardt}, {Harikane}, {Murphree}, {Stefanon}, {Wilkins}, {Amara}, {Andreon}, {Auricchio}, {Baccigalupi}, {Baldi}, {Balestra}, {Bardelli}, {Battaglia}, {Bender}, {Biviano}, {Branchini}, {Brescia}, {Brinchmann}, {Camera}, {Ca{\~n}as-Herrera}, {Capobianco}, {Carbone}, {Carretero}, {Castignani}, {Cavuoti}, {Chambers}, {Cimatti}, {Colodro-Conde}, {Congedo}, {Conversi}, {Copin}, {Courbin}, {Courtois}, {Cropper}, {Da Silva}, {Degaudenzi}, {De Lucia}, {Dole}, {Dubath}, {Duncan}, {Dupac}, {Dusini}, {Escoffier}, {Farina}, {Farinelli}, {Faustini}, {Ferriol}, {Finelli}, {Fourmanoit}, {Frailis}, {Franceschi}, {Fumana}, {Galeotta}, {George}, {Gillis}, {Giocoli}, {Gracia-Carpio}, {Grazian}, {Grupp}, {Haugan}, {Hoekstra}, {Holmes}, {Hook}, {Hormuth},
  {Hornstrup}, {Jahnke}, {Jhabvala}, {Joachimi}, {Keih{\"a}nen}, {Kermiche}, {Kiessling}, {Kubik}, {Kuijken}, {K{\"u}mmel}, {Kunz}, {Kurki-Suonio}, {Le Brun}, {Le Mignant}, {Ligori}, {Lilje}, {Lindholm}, {Lloro}, {Mainetti}, {Maino}, {Maiorano}, {Mansutti}, {Marggraf}, {Martinelli}, {Martinet}, {Marulli}, {Massey}, {Medinaceli}, {Mei}, {Mellier}, {Meneghetti}, {Merlin}, {Meylan}, {Mora}, {Moresco}, {Moscardini}, {Nakajima}, {Neissner}, {Niemi}, {Padilla}, {Paltani}, {Pasian}, {Pedersen}, {Percival}, {Pettorino}, {Pires}, {Polenta}, {Poncet}, {Popa}, {Pozzetti}, {Raison}, {Renzi}, {Rhodes}, {Riccio}, {Romelli}, {Roncarelli}, {Saglia}, {Sakr}, {Sapone}, {Sartoris}, {Schirmer}, {Schneider}, {Schrabback}, {Secroun}, {Sefusatti}, {Seidel}, {Serrano}, {Simon}, {Sirignano}, {Sirri}, {Stanco}, {Steinwagner}, {Tallada-Cresp{\'\i}}, {Taylor}, {Teplitz}, {Tereno}, {Tessore}, {Toledo-Moreo}, {Torradeflot}, {Tutusaus}, {Valenziano}, {Valiviita}, {Vassallo}, {Wang}, {Weller}, {Zamorani}, {Zerbi}, {Zucca}, {Allevato},
  {Ballardini}, {Bolzonella}, {Bozzo}, {Burigana}, {Cabanac}, {Calabrese}, {Cappi}, {Di Ferdinando}, {Escartin Vigo}, {Gabarra}, {Hartley}, {Maoli}, {Mart{\'\i}n-Fleitas}, {Matthew}, {Maturi}, {Mauri}, {Metcalf}, {Pezzotta}, {P{\"o}ntinen}, {Porciani}, {Risso}, {Scottez}, {Sereno}, {Tenti}, {Viel}, {Wiesmann}, {Akrami}, \& {Andika}}]{Euclid_Allen2026}
\textbf{Euclid Collaboration} {et~al.} 2026, \aap, 711, A25.
\bibitem[{{Faisst} {et~al.}(2026){Faisst}, {Martinez}, {Kartaltepe}, {Aravena}, {Casey}, {Silverman}, {Toft}, {Treister}, {Zavala}, {Algera}, {Battisti}, {Catone}, {Champagne}, {Cheng}, {Samir}, {Franco}, {Finner}, {Flayhart}, {Gentile}, {Gozaliasl}, {Harish}, {Jin}, {Lin}, {Liu}, {Long}, {Martin}, {Mobasher}, {Pozzi}, {Qu}, {Tian}, {Vaccari}, {Wang}, {Xu}, {Akins}, {Bagley}, {Bongiorno}, {Cooke}, {Cooper}, {Hempel-Costello}, {Delvecchio}, {Drakos}, {Enia}, {Finkelstein}, {Fujimoto}, {Gillman}, {Gomez-Guijarro}, {Gonzalez-Lopez}, {Hirschmann}, {Ilbert}, {Ito}, {Jahnke}, {Kalita}, {Kashino}, {Kokorev}, {Lai}, {Lambrides}, {Liu}, {Magdis}, {Mainieri}, {Manning}, {Maraston}, {Massey}, {McCracken}, {McKinney}, {Mercier}, {Moutard}, {Onoue}, {Paquereau}, {Renzini}, {Rhodes}, {Rich}, {Robertson}, {Rodighiero}, {Sanders}, {Scarlata}, {Scoville}, {Sheth}, {Shuntov}, {Sommovigo}, {Sparre}, {Suzuki}, {Talia}, {Trakhtenbrot}, {Valentino}, {Vanderhoof}, {Vardoulaki}, {Vijayan}, {Whitaker}, \& {Wilkins}}]{Faisst2026}
\textbf{Faisst} {et~al.} 2026, arXiv e-prints, arXiv:2609.35992.
\bibitem[{{Fanciullo} {et~al.}(2026){Fanciullo}, {Marshall}, {Kemper}, {Scicluna}, \& {Srinivasan}}]{Fanciullo2026}
\textbf{Fanciullo} {et~al.} 2026, \mnras, 551, stag1565.
\bibitem[{{Ferrara} {et~al.}(2023){Ferrara}, {Pallottini}, \& {Dayal}}]{Ferrara2023}
\textbf{Ferrara} {et~al.} 2023, \mnras, 522, 3986.
\bibitem[{{Fisher} {et~al.}(2025){Fisher}, {Bowler}, {Stefanon}, {Rowland}, {Algera}, {Aravena}, {Bouwens}, {Dayal}, {Ferrara}, {Fudamoto}, {Gulis}, {Hodge}, {Inami}, {Ormerod}, {Pallottini}, {Phillips}, {Sartorio}, {Schouws}, {Smit}, {Sommovigo}, {Stark}, \& {van der Werf}}]{Fisher2025}
\textbf{Fisher} {et~al.} 2025, \mnras, 539, 109.
\bibitem[{{Fujimoto} {et~al.}(2024{\natexlab{a}}){Fujimoto}, {Kohno}, {Ouchi}, {Oguri}, {Kokorev}, {Brammer}, {Sun}, {Gonz{\'a}lez-L{\'o}pez}, {Bauer}, {Caminha}, {Hatsukade}, {Richard}, {Smail}, {Tsujita}, {Ueda}, {Uematsu}, {Zitrin}, {Coe}, {Kneib}, {Postman}, {Umetsu}, {Lagos}, {Popping}, {Ao}, {Bradley}, {Caputi}, {Dessauges-Zavadsky}, {Egami}, {Espada}, {Ivison}, {Jauzac}, {Knudsen}, {Koekemoer}, {Magdis}, {Mahler}, {Mu{\~n}oz Arancibia}, {Rawle}, {Shimasaku}, {Toft}, {Umehata}, {Valentino}, {Wang}, \& {Wang}}]{Fujimoto2024}
\textbf{Fujimoto} {et~al.} 2024{\natexlab{a}}, \apjs, 275, 36.
\bibitem[{{Fujimoto} {et~al.}(2024{\natexlab{b}}){Fujimoto}, {Ouchi}, {Nakajima}, {Harikane}, {Isobe}, {Brammer}, {Oguri}, {Gim{\'e}nez-Arteaga}, {Heintz}, {Kokorev}, {Bauer}, {Ferrara}, {Kojima}, {Lagos}, {Laura}, {Schaerer}, {Shimasaku}, {Hatsukade}, {Kohno}, {Sun}, {Valentino}, {Watson}, {Fudamoto}, {Inoue}, {Gonz{\'a}lez-L{\'o}pez}, {Koekemoer}, {Knudsen}, {Lee}, {Magdis}, {Richard}, {Strait}, {Sugahara}, {Tamura}, {Toft}, {Umehata}, \& {Walth}}]{Fujimoto2024_ID4590}
\textbf{Fujimoto} {et~al.} 2024{\natexlab{b}}, \apj, 964, 146.
\bibitem[{{Galliano} {et~al.}(2021){Galliano}, {Nersesian}, {Bianchi}, {De Looze}, {Roychowdhury}, {Baes}, {Casasola}, {Cassar{\'a}}, {Dobbels}, {Fritz}, {Galametz}, {Jones}, {Madden}, {Mosenkov}, {Xilouris}, \& {Ysard}}]{Galliano2021}
\textbf{Galliano} {et~al.} 2021, \aap, 649, A18.
\bibitem[{{Geesink} {et~al.}(2026){Geesink}, {Popping}, {Boogaard}, {Langan}, {Pope}, {Popescu}, {Shivaei}, {Solimano}, {Kaasinen}, {Mobasher}, {Narayanan}, {Parente}, {Reddy}, \& {Sanders}}]{Geesink2026}
\textbf{Geesink} {et~al.} 2026, arXiv e-prints, arXiv:2609.20926.
\bibitem[{{Ginolfi} {et~al.}(2020){Ginolfi}, {Jones}, {B{\'e}thermin}, {Fudamoto}, {Loiacono}, {Fujimoto}, {Le F{\'e}vre}, {Faisst}, {Schaerer}, {Cassata}, {Silverman}, {Yan}, {Capak}, {Bardelli}, {Boquien}, {Carraro}, {Dessauges-Zavadsky}, {Giavalisco}, {Gruppioni}, {Ibar}, {Khusanova}, {Lemaux}, {Maiolino}, {Narayanan}, {Oesch}, {Pozzi}, {Rodighiero}, {Talia}, {Toft}, {Vallini}, {Vergani}, \& {Zamorani}}]{Ginolfi2020}
\textbf{Ginolfi} {et~al.} 2020, \aap, 633, A90.
\bibitem[{{Glenn} {et~al.}(2025){Glenn}, {Meixner}, {Bradford}, {Pontoppidan}, {Pope}, {Kataria}, {Rocca}, {Luthman}, {Armus}, {Baselmans}, {Battersby}, {Bollato}, {Burgarella}, {Chen}, {Ciesla}, {Day}, {Di Giorgio}, {Dipirro}, {Dowell}, {Echternach}, {Essinger-Hileman}, {Foote}, {Gruppioni}, {Hensley}, {Henning}, {Jellema}, {Johnson}, {Kogut}, {Krause}, {McGuire}, {Mills}, {Moullet}, {Rodgers}, {Sauvage}, {Smith}, {Somerville}, {Staguhn}, {Stevenson}, {Tucker}, {Unwin}, {Ziemer}, {Cannella}, \& {Dissly}}]{Glenn2025}
\textbf{Glenn} {et~al.} 2025, Journal of Astronomical Telescopes, Instruments, and Systems, 11, 031628.
\bibitem[{{Harikane} {et~al.}(2023){Harikane}, {Ouchi}, {Oguri}, {Ono}, {Nakajima}, {Isobe}, {Umeda}, {Mawatari}, \& {Zhang}}]{Harikane2023}
\textbf{Harikane} {et~al.} 2023, \apjs, 265, 5.
\bibitem[{{Hashimoto} {et~al.}(2018){Hashimoto}, {Laporte}, {Mawatari}, {Ellis}, {Inoue}, {Zackrisson}, {Roberts-Borsani}, {Zheng}, {Tamura}, {Bauer}, {Fletcher}, {Harikane}, {Hatsukade}, {Hayatsu}, {Matsuda}, {Matsuo}, {Okamoto}, {Ouchi}, {Pell{\'o}}, {Rydberg}, {Shimizu}, {Taniguchi}, {Umehata}, \& {Yoshida}}]{Hashimoto2019}
\textbf{Hashimoto} {et~al.} 2018, \nat, 557, 392.
\bibitem[{{Heintz} {et~al.}(2025){Heintz}, {Pollock}, {Witstok}, {Carniani}, {Hainline}, {D'Eugenio}, {Terp}, {Saxena}, \& {Watson}}]{Heintz2025}
\textbf{Heintz} {et~al.} 2025, \apjl, 987, L2.
\bibitem[{{Hirashita}(2012)}]{Hirashita2012}
\textbf{Hirashita}, H. 2012, \mnras, 422, 1263.
\bibitem[{{Hirashita} \& {Aoyama}(2019)}]{HirashitaAoyama2019}
\textbf{Hirashita} {et~al.} 2019, \mnras, 482, 2555.
\bibitem[{{Inami} {et~al.}(2022){Inami}, {Algera}, {Schouws}, {Sommovigo}, {Bouwens}, {Smit}, {Stefanon}, {Bowler}, {Endsley}, {Ferrara}, {Oesch}, {Stark}, {Aravena}, {Barrufet}, {da Cunha}, {Dayal}, {De Looze}, {Fudamoto}, {Gonzalez}, {Graziani}, {Hodge}, {Hygate}, {Nanayakkara}, {Pallottini}, {Riechers}, {Schneider}, {Topping}, \& {van der Werf}}]{Inami2022}
\textbf{Inami} {et~al.} 2022, \mnras, 515, 3126.
\bibitem[{{Inoue} {et~al.}(2022){Inoue}, {Kaneda}, {Yamada}, {Harikane}, {Ishihara}, {Kodama}, {Komiyama}, {Moriya}, {Motohara}, {Nomura}, {Ouchi}, {Oyabu}, {Suzuki}, {Wada}, \& {Yamamura}}]{Inoue2022}
\textbf{Inoue} {et~al.} 2022, in SPIE 2022: Optical, Infrared, and Millimeter Wave, Vol. 12180, 121801I.
\bibitem[{{K{\"o}hler} {et~al.}(2012){K{\"o}hler}, {Stepnik}, {Jones}, {Guillet}, {Abergel}, {Ristorcelli}, \& {Bernard}}]{Kohler2012}
\textbf{K{\"o}hler} {et~al.} 2012, \aap, 548, A61.
\bibitem[{{Liu} {et~al.}(2026){Liu}, {Dunlop}, {McLure}, {McLeod}, {Barrufet}, {Carnall}, {Begley}, {P{\'e}rez-Gonz{\'a}lez}, {Donnan}, {Ellis}, {Grogin}, {Magee}, {Illingworth}, {Cullen}, {Stevenson}, {Koekemoer}, {Fontana}, \& {Bowler}}]{Liu2026}
\textbf{Liu} {et~al.} 2026, \mnras, 545, staf1961.
\bibitem[{{Markov} {et~al.}(2025){Markov}, {Gallerani}, {Ferrara}, {Pallottini}, {Parlanti}, {Mascia}, {Sommovigo}, \& {Kohandel}}]{Markov2025}
\textbf{Markov} {et~al.} 2025, Nature Astronomy, 9, 458.
\bibitem[{{Matsumoto} {et~al.}(2026){Matsumoto}, {Sommovigo}, {Gebek}, {Nagamine}, {Nersesian}, {Baes}, {De Looze}, {van der Wel}, {Somerville}, {Romano}, \& {Cochrane}}]{Matsumoto2025}
\textbf{Matsumoto} {et~al.} 2026, \aap, 705, A75.
\bibitem[{{Maud} {et~al.}(2023){Maud}, {Asaki}, {Nagai}, {Tsukui}, {Hirota}, {Fomalont}, {Dent}, {Takahashi}, \& {Phillips}}]{Maud2023}
\textbf{Maud} {et~al.} 2023, \apjs, 267, 24.
\bibitem[{{Micha{\l}owski}(2015)}]{Michalowski2015}
\textbf{Micha{\l}owski}, M.~J. 2015, \aap, 577, A80.
\bibitem[{{Mitsuhashi} {et~al.}(2026{\natexlab{a}}){Mitsuhashi}, {Suess}, {Leja}, {Dayal}, {Feldmann}, {Fujimoto}, {Katz}, {Nanayakkara}, {Narayanan}, {Price}, {Weaver}, {Williams}, {Labbe}, {Bezanson}, {Atek}, {Brammer}, {Cutler}, {Furtak}, {Pan}, {Wang}, {Whitaker}, \& {The Uncover/Megascience Team}}]{Mitsuhashi2026b}
\textbf{Mitsuhashi} {et~al.} 2026{\natexlab{a}}, \apj, 1005, 108.
\bibitem[{{Mitsuhashi} {et~al.}(2026{\natexlab{b}}){Mitsuhashi}, {Zavala}, {Bakx}, {Inoue}, {Castellano}, {Calabr{\`o}}, {Casey}, {Franco}, {Hatsukade}, {Hathi}, {Ikeda}, {Koekemoer}, {Kartaltepe}, {Knudsen}, {Santini}, {Saito}, {Terlevich}, {Terlevich}, \& {Yung}}]{Mitsuhashi2026}
\textbf{Mitsuhashi} {et~al.} 2026{\natexlab{b}}, \apj, 1000, 159.
\bibitem[{{Nanni} {et~al.}(2014){Nanni}, {Bressan}, {Marigo}, \& {Girardi}}]{Nanni2014}
\textbf{Nanni} {et~al.} 2014, \mnras, 438, 2328.
\bibitem[{{Narayanan} {et~al.}(2026){Narayanan}, {Torrey}, {Stark}, {Chisholm}, {Finkelstein}, {Garcia}, {Kelley-Derzon}, {Marinacci}, {Sales}, {Savitch}, {Vogelsberger}, \& {Zimmerman}}]{Narayanan2026}
\textbf{Narayanan} {et~al.} 2026, The Open Journal of Astrophysics, 9, 59986.
\bibitem[{{Ocvirk} {et~al.}(2025){Ocvirk}, {Lewis}, {Conaboy}, {Dubois}, {Bethermin}, {Sorce}, {Aubert}, {Shapiro}, {Dawoodbhoy}, {Lee}, {Teyssier}, {Yepes}, {Gottl{\"o}ber}, {Iliev}, {Ahn}, {Park}, \& {Palanque}}]{Ocvirk2025}
\textbf{Ocvirk} {et~al.} 2025, \aap, 703, A98.
\bibitem[{{Osman} {et~al.}(2026){Osman}, {De Lucia}, {Fontanot}, {Xie}, \& {Hirschmann}}]{Osman2026}
\textbf{Osman} {et~al.} 2026, \aap, 710, A306.
\bibitem[{{Otaki} {et~al.}(2026){Otaki}, {Schneider}, {Bianchi}, {Witstok}, {Graziani}, {Limongi}, \& {Maiolino}}]{Otaki2026}
\textbf{Otaki} {et~al.} 2026, arXiv e-prints, arXiv:2604.26750.
\bibitem[{{Parente} {et~al.}(2026{\natexlab{a}}){Parente}, {Narayanan}, \& {Torrey}}]{Parente2026_grainsizes}
\textbf{Parente} {et~al.} 2026{\natexlab{a}}, The Open Journal of Astrophysics, 9, 66984.
\bibitem[{{Parente} {et~al.}(2026{\natexlab{b}}){Parente}, {Salvestrini}, {Granato}, {Narayanan}, {Tripodi}, {Bianchi}, {Bischetti}, {Feruglio}, {Fiore}, \& {Silva}}]{Parente2026}
\textbf{Parente} {et~al.} 2026{\natexlab{b}}, arXiv e-prints, arXiv:2603.04505.
\bibitem[{{Pollock} {et~al.}(2026){Pollock}, {Heintz}, {Witstok}, {Gottumukkala}, {Brammer}, {Bose}, {Cameron}, {Dayal}, {van Dokkum}, {Fynbo}, {Gelli}, {Hayes}, {Inoue}, {Lagos}, {Laursen}, {Meyer}, {Naidu}, {Oesch}, {Rowland}, {Tanvir}, {Tacchella}, {Terp}, {Valentino}, {Walter}, {Weaver}, \& {Witten}}]{Pollock2026}
\textbf{Pollock} {et~al.} 2026, \aap, 711, A298.
\bibitem[{{Popping} {et~al.}(2017){Popping}, {Somerville}, \& {Galametz}}]{Popping2017}
\textbf{Popping} {et~al.} 2017, \mnras, 471, 3152.
\bibitem[{{Prandoni} {et~al.}(2026){Prandoni}, {Sargent}, {Bonato}, {Lal}, {Mousavi-Sadr}, {Ghasemi-Nodehi}, {Seymour}, {Tabatabaei}, {De Zotti}, {Baronchelli}, {Liuzzo}, {Marchili}, {Massardi}, \& {Paladino}}]{Prandoni2026_ska}
\textbf{Prandoni} {et~al.} 2026, Advancing Astrophysics with the SKA I, 68.
\bibitem[{{R{\'e}my-Ruyer} {et~al.}(2014){R{\'e}my-Ruyer}, {Madden}, {Galliano}, {Galametz}, {Takeuchi}, {Asano}, {Zhukovska}, {Lebouteiller}, {Cormier}, {Jones}, {Bocchio}, {Baes}, {Bendo}, {Boquien}, {Boselli}, {DeLooze}, {Doublier-Pritchard}, {Hughes}, {Karczewski}, \& {Spinoglio}}]{RemyRuyer2014}
\textbf{R{\'e}my-Ruyer} {et~al.} 2014, \aap, 563, A31.
\bibitem[{{Rodighiero} {et~al.}(2026){Rodighiero}, {Ferrara}, {Catone}, {Napolitano}, {Cassata}, {Gandolfi}, {Merlin}, {Grazian}, {Renzini}, {Bisigello}, {Castellano}, {P{\'e}rez-Gonz{\'a}lez}, {P{\'e}rez-D{\'\i}az}, {Iani}, {Gruppioni}, {Finkelstein}, {Koekemoer}, {Bianchetti}, \& {Sinigaglia}}]{Rodighiero2026}
\textbf{Rodighiero} {et~al.} 2026, arXiv e-prints, arXiv:2603.15841.
\bibitem[{{Rowland} {et~al.}(2026{\natexlab{a}}){Rowland}, {Algera}, {Hodge}, {Stefanon}, {Bouwens}, {Aravena}, {Astles}, {Cescon}, {da Cunha}, {de Looze}, {Ferrara}, {Fisher}, {Fudamoto}, {Herard-Demanche}, {Inami}, {Kohandel}, {Komarova}, {Laza Ramos}, {Nanayakkara}, {Ormerod}, {Pallottini}, {Phillips}, {Schouws}, {Sharda}, {Smit}, \& {van der Werf}}]{Rowland2026}
\textbf{Rowland} {et~al.} 2026{\natexlab{a}}, arXiv e-prints, arXiv:2608.25982.
\bibitem[{{Rowland} {et~al.}(2026{\natexlab{b}}){Rowland}, {Heintz}, {Algera}, {Stefanon}, {Hodge}, {Bouwens}, {Aravena}, {Astles}, {da Cunha}, {Dayal}, {Ferrara}, {Fisher}, {Gonz{\'a}lez}, {Inami}, {Komarova}, {de Looze}, {Nanayakkara}, {Ormerod}, {Pallottini}, {Pollock}, {Smit}, {van der Werf}, \& {Witstok}}]{Rowland2026_dla}
\textbf{Rowland} {et~al.} 2026{\natexlab{b}}, \aap, 710, A374.
\bibitem[{{Rowland} {et~al.}(2024){Rowland}, {Hodge}, {Bouwens}, {Mancera Pi{\~n}a}, {Hygate}, {Algera}, {Aravena}, {Bowler}, {da Cunha}, {Dayal}, {Ferrara}, {Herard-Demanche}, {Inami}, {van Leeuwen}, {de Looze}, {Oesch}, {Pallottini}, {Phillips}, {Rybak}, {Schouws}, {Smit}, {Sommovigo}, {Stefanon}, \& {van der Werf}}]{Rowland2024}
\textbf{Rowland} {et~al.} 2024, \mnras, 535, 2068.
\bibitem[{{Schneider} \& {Maiolino}(2024)}]{SchneiderMaiolino2024}
\textbf{Schneider} {et~al.} 2024, \aapr, 32, 2.
\bibitem[{{Schouws} {et~al.}(2025){Schouws}, {Bouwens}, {Ormerod}, {Smit}, {Algera}, {Sommovigo}, {Hodge}, {Ferrara}, {Oesch}, {Rowland}, {van Leeuwen}, {Stefanon}, {Herard-Demanche}, {Fudamoto}, {R{\"o}ttgering}, \& {van der Werf}}]{Schouws2025}
\textbf{Schouws} {et~al.} 2025, \apj, 988, 19.
\bibitem[{{Schreiber} {et~al.}(2018){Schreiber}, {Elbaz}, {Pannella}, {Ciesla}, {Wang}, \& {Franco}}]{Schreiber2018}
\textbf{Schreiber} {et~al.} 2018, \aap, 609, A30.
\bibitem[{{Shivaei} {et~al.}(2026){Shivaei}, {Naidu}, {Rodr{\'\i}guez Montero}, {Matsumoto}, {Leja}, {Matthee}, {Johnson}, {Oesch}, {Chevallard}, {Adamo}, {Bodansky}, {Bunker}, {Covelo Paz}, {Di Cesare}, {Egami}, {Furtak}, {Heintz}, {Kramarenko}, {Meyer}, {Reddy}, {Rinaldi}, {Tacchella}, {Torralba}, {Witstok}, {Wozniak}, \& {Xiao}}]{Shivaei2026}
\textbf{Shivaei} {et~al.} 2026, \aap, 712, A132.
\bibitem[{{Sommovigo} \& {Algera}(2025)}]{SommovigoAlgera2025}
\textbf{Sommovigo} {et~al.} 2025, \mnras, 540, 3693.
\bibitem[{{Sommovigo} {et~al.}(2022){Sommovigo}, {Ferrara}, {Pallottini}, {Dayal}, {Bouwens}, {Smit}, {da Cunha}, {De Looze}, {Bowler}, {Hodge}, {Inami}, {Oesch}, {Endsley}, {Gonzalez}, {Schouws}, {Stark}, {Stefanon}, {Aravena}, {Graziani}, {Riechers}, {Schneider}, {van der Werf}, {Algera}, {Barrufet}, {Fudamoto}, {Hygate}, {Labb{\'e}}, {Li}, {Nanayakkara}, \& {Topping}}]{Sommovigo2022}
\textbf{Sommovigo} {et~al.} 2022, MNRAS, 513, 3122.
\bibitem[{{Sommovigo} {et~al.}(2026){Sommovigo}, {Lancaster}, {Menon}, {O'Leary}, {Somerville}, \& {Bryan}}]{Sommovigo2026}
\textbf{Sommovigo} {et~al.} 2026, \apj, 1003, 170.
\bibitem[{{Tailor} {et~al.}(2026){Tailor}, {Casasola}, {Calura}, {Pozzi}, {Fritz}, {Lara-L{\'o}pez}, {Palla}, {Bianchi}, {{\'A}lvarez-C{\'o}rdoba}, {Bonato}, {De Rossi}, {Dib}, {Gardu{\~n}o}, {Giannetti}, {L{\'o}pez Cruz}, {Puebla}, {Paspaliaris}, {Pilyugin}, {Val{\'e}}, \& {Valerdi}}]{Tailor2026}
\textbf{Tailor} {et~al.} 2026, arXiv e-prints, arXiv:2609.04320.
\bibitem[{{Tamura} {et~al.}(2023){Tamura}, {C. Bakx}, {Inoue}, {Hashimoto}, {Tokuoka}, {Imamura}, {Hatsukade}, {Lee}, {Moriwaki}, {Okamoto}, {Ota}, {Umehata}, {Yoshida}, {Zackrisson}, {Hagimoto}, {Matsuo}, {Shimizu}, {Sugahara}, \& {Takeuchi}}]{Tamura2023}
\textbf{Tamura} {et~al.} 2023, \apj, 952, 9.
\bibitem[{{Tamura} {et~al.}(2019){Tamura}, {Mawatari}, {Hashimoto}, {Inoue}, {Zackrisson}, {Christensen}, {Binggeli}, {Matsuda}, {Matsuo}, {Takeuchi}, {Asano}, {Sunaga}, {Shimizu}, {Okamoto}, {Yoshida}, {Lee}, {Shibuya}, {Taniguchi}, {Umehata}, {Hatsukade}, {Kohno}, \& {Ota}}]{Tamura2019}
\textbf{Tamura} {et~al.} 2019, \apj, 874, 27.
\bibitem[{{Todini} \& {Ferrara}(2001)}]{TodiniFerrara2001}
\textbf{Todini} {et~al.} 2001, \mnras, 325, 726.
\bibitem[{{Toyouchi} {et~al.}(2025){Toyouchi}, {Yajima}, {Ferrara}, \& {Nagamine}}]{Toyouchi2025}
\textbf{Toyouchi} {et~al.} 2025, \mnras, 541, 3606.
\bibitem[{{Traina} {et~al.}(2026){Traina}, {Gruppioni}, {Delvecchio}, {Magnelli}, {Calura}, {Bisigello}, {Feltre}, {Vallini}, {De Lucia}, {Fontanot}, \& et~al.}]{Traina2026}
\textbf{Traina} {et~al.} 2026, \aap, 705, A255.
\bibitem[{{Tripodi} {et~al.}(2024){Tripodi}, {Feruglio}, {Fiore}, {Zappacosta}, {Piconcelli}, {Bischetti}, {Bongiorno}, {Carniani}, {Civano}, {Chen}, {Cristiani}, {Cupani}, {Di Mascia}, {D'Odorico}, {Fan}, {Ferrara}, {Gallerani}, {Ginolfi}, {Maiolino}, {Mainieri}, {Marconi}, {Saccheo}, {Salvestrini}, {Tortosa}, \& {Valiante}}]{Tripodi2024}
\textbf{Tripodi} {et~al.} 2024, \aap, 689, A220.
\bibitem[{{Umehata} {et~al.}(2025){Umehata}, {Tamura}, {Fudamoto}, {Nakazato}, {Ceverino}, {Yoshida}, {Inoue}, {Ikeda}, {Sugahara}, {Inui}, {Arribas}, {Bakx}, {Hagimoto}, {Hashimoto}, {Colina}, {Ren}, {Osone}, {Crespo G{\'o}mez}, {Mawatari}, \& {{\'A}lvarez-M{\'a}rquez}}]{Umehata2025}
\textbf{Umehata} {et~al.} 2025, \apjl, 993, L50.
\bibitem[{{van Kampen} {et~al.}(2024){van Kampen}, {Bakx}, {De Breuck}, {Chen}, {Dannerbauer}, {Magnelli}, {Montenegro-Montes}, {Okumura}, {Pu}, {Rybak}, {Saintonge}, {Cicone}, {Hatziminaoglou}, {Hilhorst}, {Klaassen}, {Lee}, {Lovell}, {Lundgren}, {Di Mascolo}, {Mroczkowski}, {Sommovigo}, {Booth}, {Cordiner}, {Ivison}, {Johnstone}, {Liu}, {Maccarone}, {Smith}, {Thelen}, \& {Wedemeyer}}]{vanKampen2024}
\textbf{van Kampen} {et~al.} 2024, Open Research Europe, 4, 122.
\bibitem[{{Viero} {et~al.}(2022){Viero}, {Sun}, {Chung}, {Moncelsi}, \& {Condon}}]{Viero2022}
\textbf{Viero} {et~al.} 2022, \mnras, 516, L30.
\bibitem[{{Vijayan} {et~al.}(2024){Vijayan}, {Thomas}, {Lovell}, {Wilkins}, {Greve}, {Irodotou}, {Roper}, \& {Seeyave}}]{Vijayan2024}
\textbf{Vijayan} {et~al.} 2024, \mnras, 527, 7337.
\bibitem[{{Vijayan} {et~al.}(2026){Vijayan}, {Trayford}, {Schaye}, {Ploeckinger}, {Gebek}, {Andreadis}, {Baes}, {Ben{\'\i}tez-Llambay}, {Chaikin}, {Frenk}, {Hu{\v{s}}ko}, {McGibbon}, {Richings}, \& {Schaller}}]{Vijayan2026}
\textbf{Vijayan} {et~al.} 2026, arXiv e-prints, arXiv:2607.26058.
\bibitem[{{Walter} {et~al.}(2016){Walter}, {Decarli}, {Aravena}, {Carilli}, {Bouwens}, {da Cunha}, {Daddi}, {Ivison}, {Riechers}, {Smail}, {Swinbank}, {Weiss}, {Anguita}, {Assef}, {Bacon}, {Bauer}, {Bell}, {Bertoldi}, {Chapman}, {Colina}, {Cortes}, {Cox}, {Dickinson}, {Elbaz}, {G{\'o}nzalez-L{\'o}pez}, {Ibar}, {Inami}, {Infante}, {Hodge}, {Karim}, {Le Fevre}, {Magnelli}, {Neri}, {Oesch}, {Ota}, {Popping}, {Rix}, {Sargent}, {Sheth}, {van der Wel}, {van der Werf}, \& {Wagg}}]{Walter2016}
\textbf{Walter} {et~al.} 2016, \apj, 833, 67.
\bibitem[{{Wang} {et~al.}(2026){Wang}, {Wagg}, {Sargent}, {Aravena}, {Pandey-Pommier}, {da Cunha}, {Murphy}, {Riechers}, \& {Hodge}}]{Wang2026_ska}
\textbf{Wang} {et~al.} 2026, Advancing Astrophysics with the SKA I, 67.
\bibitem[{{Watson} {et~al.}(2015){Watson}, {Christensen}, {Knudsen}, {Richard}, {Gallazzi}, \& {Micha{\l}owski}}]{Watson2015}
\textbf{Watson} {et~al.} 2015, \nat, 519, 327.
\bibitem[{{Weibel} {et~al.}(2024){Weibel}, {Oesch}, {Barrufet}, {Gottumukkala}, {Ellis}, {Santini}, {Weaver}, {Allen}, {Bouwens}, {Bowler}, {Brammer}, {Carnall}, {Cullen}, {Dayal}, {Dickinson}, {Donnan}, {Dunlop}, {Giavalisco}, {Grogin}, {Illingworth}, {Koekemoer}, {Labbe}, {Marchesini}, {McLeod}, {McLure}, {Naidu}, {P{\'e}rez-Gonz{\'a}lez}, {Shuntov}, {Stefanon}, {Toft}, \& {Xiao}}]{Weibel2024}
\textbf{Weibel} {et~al.} 2024, \mnras, 533, 1808.
\bibitem[{{Weibel} {et~al.}(2026){Weibel}, {Oesch}, {Williams}, {Jespersen}, {Shuntov}, {Whitaker}, {Atek}, {Bezanson}, {Brammer}, {Chemerynska}, {Cloonan}, {Dayal}, {Furtak}, {Hutter}, {Ji}, {Maseda}, \& {Xiao}}]{Weibel2026}
\textbf{Weibel} {et~al.} 2026, \apj, 1002, 136.
\bibitem[{{Witstok} {et~al.}(2023){Witstok}, {Jones}, {Maiolino}, {Smit}, \& {Schneider}}]{Witstok2023}
\textbf{Witstok} {et~al.} 2023, \mnras, 523, 3119.
\bibitem[{{Witstok} {et~al.}(2022){Witstok}, {Smit}, {Maiolino}, {Kumari}, {Aravena}, {Boogaard}, {Bouwens}, {Carniani}, {Hodge}, {Jones}, {Stefanon}, {van der Werf}, \& {Schouws}}]{Witstok2022}
\textbf{Witstok} {et~al.} 2022, \mnras, 515, 1751.
\bibitem[{{Zanella} {et~al.}(2018){Zanella}, {Daddi}, {Magdis}, {Diaz Santos}, {Cormier}, {Liu}, {Cibinel}, {Gobat}, {Dickinson}, {Sargent}, {Popping}, {Madden}, {Bethermin}, {Hughes}, {Valentino}, {Rujopakarn}, {Pannella}, {Bournaud}, {Walter}, {Wang}, {Elbaz}, \& {Coogan}}]{Zanella2018}
\textbf{Zanella} {et~al.} 2018, \mnras, 481, 1976.
\bibitem[{{Zavala} {et~al.}(2024){Zavala}, {Bakx}, {Mitsuhashi}, {Castellano}, {Calabro}, {Akins}, {Buat}, {Casey}, {Fernandez-Arenas}, {Franco}, {Fontana}, {Hatsukade}, {Ho}, {Ikeda}, {Kartaltepe}, {Koekemoer}, {McKinney}, {Napolitano}, {P{\'e}rez-Gonz{\'a}lez}, {Santini}, {Serjeant}, {Terlevich}, {Terlevich}, \& {Yung}}]{Zavala2024}
\textbf{Zavala} {et~al.} 2024, \apjl, 977, L9.
\bibitem[{{Zavala} {et~al.}(2021){Zavala}, {Casey}, {Manning}, {Aravena}, {Bethermin}, {Caputi}, {Clements}, {Cunha}, {Drew}, {Finkelstein}, {Fujimoto}, {Hayward}, {Hodge}, {Kartaltepe}, {Knudsen}, {Koekemoer}, {Long}, {Magdis}, {Man}, {Popping}, {Sanders}, {Scoville}, {Sheth}, {Staguhn}, {Toft}, {Treister}, {Vieira}, \& {Yun}}]{Zavala2021}
\textbf{Zavala} {et~al.} 2021, \apj, 909, 165.
\bibitem[{{Zavala} {et~al.}(2025){Zavala}, {Castellano}, {Akins}, {Bakx}, {Burgarella}, {Casey}, {Ch{\'a}vez Ortiz}, {Dickinson}, {Finkelstein}, {Mitsuhashi}, {Nakajima}, {P{\'e}rez-Gonz{\'a}lez}, {Arrabal Haro}, {Bergamini}, {Buat}, {Backhaus}, {Calabr{\`o}}, {Cleri}, {Fern{\'a}ndez-Arenas}, {Fontana}, {Franco}, {Grillo}, {Giavalisco}, {Grogin}, {Hathi}, {Hirschmann}, {Ikeda}, {Jung}, {Kartaltepe}, {Koekemoer}, {Larson}, {McKinney}, {Papovich}, {Rosati}, {Saito}, {Santini}, {Terlevich}, {Terlevich}, {Treu}, \& {Yung}}]{Zavala2025}
\textbf{Zavala} {et~al.} 2025, Nature Astronomy, 9, 155.
\end{thebibliography}

\clearpage 

{\footnotesize{\vspace{1em}\noindent
$^{1}$Institute of Astronomy and Astrophysics, Academia Sinica, 11F of Astronomy-Mathematics Building, No.1, Sec. 4, Roosevelt Rd, Taipei 106319, Taiwan, R.O.C., \\
$^{2}$Department of Physics and Astronomy, Chalmers University of Technology, SE-412 96 Gothenburg, Sweden, \\
$^{3}$Instituto de Estudios Astrof\'{i}sicos, Facultad de Ingenier\'{i}a y Ciencias, Universidad Diego Portales, Avenida Ej\'{e}rcito Libertador 441, Santiago, Chile, \\
$^{4}$Leiden Observatory, Leiden University, P.O. Box 9513, 2300 RA Leiden, The Netherlands, \\
$^{5}$Kapteyn Astronomical Institute, University of Groningen, P.O. Box 800, 9700AV Groningen, The Netherlands, \\
$^{6}$Scuola Normale Superiore, Piazza dei Cavalieri 7, I-56126 Pisa, Italy, \\
$^{7}$Sorbonne Université, UPMC Paris 6 and CNRS, UMR 7095, Institut d’Astrophysique de Paris, 98b bd.\ Arago, 75014 Paris, France, \\
$^{8}$International Centre for Radio Astronomy Research, The University of Western Australia, 35 Stirling Hwy, Crawley WA 6009, Australia, \\
$^{9}$Department of Astronomy, University of Geneva, Chemin Pegasi 51, 1290 Versoix, Switzerland, \\
$^{10}$Foundation for Research and Technology--Hellas (FORTH), Heraklion, GR-70013, Greece, \\
$^{11}$IPAC, California Institute of Technology, 1200 E. California Blvd. Pasadena, CA 91125, USA, \\
$^{12}$Center for Frontier Science, Chiba University, 1-33 Yayoi-cho, Inage-ku, Chiba 263-8522, Japan, \\
$^{13}$INAF, Osservatorio di Astrofisica e Scienza dello Spazio, Via P. Gobetti, 93/3, 40129 Bologna, Italy, \\ 
$^{14}$Centro de Astrobiología (CAB/CSIC-INTA), Ctra. de Ajalvir km 4, Torrejón de Ardoz, 28850, Madrid, Spain, \\ 
$^{15}$Hiroshima Astrophysical Science Center, Hiroshima University, 1-3-1 Kagamiyama, Higashi-Hiroshima, Hiroshima 739-8526, Japan, \\
$^{16}$Instituto de Astronomía y Ciencias Planetarias, Universidad de Atacama, Copayapu 485, 1531772, Copiapó, Chile, \\
$^{17}$SISSA, Via Bonomea 265, 34136 Trieste, Italy, \\
$^{18}$Purple Mountain Observatory, Chinese Academy of Sciences, 10 Yuanhua Road, Nanjing 210023, China, \\
$^{19}$State Key Laboratory of Radio Astronomy and Technology, Purple Mountain Observatory, Chinese Academy of Sciences, 10 Yuanhua Road, Nanjing 210023, China, \\
$^{20}$INAF, Osservatorio Astronomico di Trieste, via Tiepolo 11, I-34131, Trieste, Italy, \\
$^{21}$Institute for Fundamental Physics of the Universe (IFPU), Via Beirut 2, I-34151 Trieste, Italy, \\
$^{22}$Department of Astronomy, University of Michigan, 1085 South University Avenue, Ann Arbor, MI 48109, USA, \\
$^{23}$University of Khartoum – Department of Physics, Al–Gamaa Ave, 11115 Khartoum, Sudan, \\
$^{24}$Dipartimento di Fisica “Enrico Fermi”, Universitá di Pisa, Largo Bruno Pontecorvo 3, Pisa I-56127, Italy, \\
$^{25}$ESO, Karl-Schwarzschild-Strasse 2, D-85748 Garching, Germany, \\
$^{26}$Department of Astronomy, Columbia University, New York, NY 10027, USA, \\
$^{27}$Center for Computational Astrophysics, Flatiron Institute, 162 Fifth Avenue, New York, NY 10010, USA, \\
$^{28}$Department of Physics and Astronomy and George P. and Cynthia Woods Mitchell Institute for Fundamental Physics and Astronomy, Texas A\&M University, 4242 TAMU, College Station, TX 77843-4242, USA, \\
$^{29}$Physics Department, Lancaster University, Lancaster LA1 4YB, UK, \\
$^{30}$Cosmic Dawn Center (DAWN), Copenhagen, Denmark, \\
$^{31}$Niels Bohr Institute, University of Copenhagen, Jagtvej 128, DK-2200, Copenhagen, Denmark
}}

\end{document}